\documentclass[useAMS,usenatbib]{mn2e}
\usepackage{epsf}
\usepackage{amssymb}
\usepackage[usenames]{color}

\def\plotone#1{\centering \leavevmode
\epsfxsize=\columnwidth \epsfbox{#1}}
\def\plottwo#1#2{\centering \leavevmode
\epsfxsize=.99\columnwidth \epsfbox{#1} \hfil
\epsfxsize=.99\columnwidth \epsfbox{#2}}

\def\deg{$^{\circ}$}

\title[Non-thermal pressure in M87 and NGC1399]{Measuring the non-thermal pressure in early type galaxy atmospheres:
A comparison of X-ray and optical potential profiles in M87 and NGC1399}

\author[Churazov et al.]{E.~Churazov,$^{1,2}$ W.~Forman,$^{3}$ A.~Vikhlinin,$^{3,2}$
S.~Tremaine,$^{4}$ O.~Gerhard,$^{5}$ C.~Jones$^{3}$ \newauthor \\
$^1$ Max-Planck-Institut f\"ur Astrophysik, Karl-Schwarzschild-Strasse 1, 85741
Garching, Germany\\
$^2$ Space Research Institute (IKI), Profsoyuznaya 84/32, Moscow 117810, 
Russia\\
$^3$ Harvard-Smithsonian Center for Astrophysics, 60 Garden St.,
Cambridge, MA 02138, USA \\
$^4$ Institute for Advanced Study, Einstein Dr., Princeton, NJ 08540, USA \\
$^5$ MPI f\"{u}r Extraterrestrische Physik, P.O.Box 1603, 85740
Garching, Germany\\
}

\begin{document}

\pagerange{\pageref{firstpage}--\pageref{lastpage}}
\pubyear{2001}

\maketitle

\label{firstpage}
\begin{abstract}
We compare the gravitational potential profiles of the elliptical
galaxies NGC 4486 (M87) and NGC 1399 (the central galaxy in the Fornax
cluster) derived from X-ray and optical data. This comparison suggests
that the combined contribution of cosmic rays, magnetic fields and
micro-turbulence to the pressure is $\sim$10\% of the gas thermal
pressure in the cores of NGC 1399 and M87, although the uncertainties
in our model assumptions (e.g., spherical symmetry) are sufficiently
large that the contribution could be consistent with zero.  In the
absence of any other form of non-thermal pressure support, these upper
bounds translate into upper limits on the magnetic field of
$\sim$10-20 $\mu$G at a distance of $1'$-$2'$ from the centers of
NGC1399 and M87.  We show that these results are consistent with the
current paradigm of cool cluster cores, based on the assumption that
AGN activity regulates the thermal state of the gas by injecting
energy into the intra-cluster medium.  The limit of $\sim$10-20\% on
the energy density in the form of relativistic protons applies not
only to the current state of the gas, but essentially to the entire
history of the intra-cluster medium, provided that cosmic ray protons
evolve adiabatically and that their spatial diffusion is suppressed.

\end{abstract}


%

\sloppypar

\section{Introduction}

Both optical and X-ray data are often used to determine the
distribution of the gravitating mass in galaxies.  In
analysing optical stellar-kinematic data,
stars are treated as collisionless particles and the mass distribution
is obtained either from the Jeans equations and some assumption about
the anisotropy of stellar orbits (Binney \& Tremaine 2008;
Lokas \& Mamon 2003), or by finding the potentials in which 
a distribution of orbits reproduces
the observed surface brightness profile and stellar kinematics
(e.g., Kronawitter et al. 2000; Thomas et al. 2007). In X-rays,
hydrostatic equilibrium is usually assumed for the gaseous atmosphere
and the observed density and temperature distributions are then used
to evaluate the distribution of gravitating mass (Mathews 1978;
Forman, Jones \& Tucker 1985; Fukazawa et al. 2006; Humphrey et
al. 2006).  Ideally both methods should yield identical results, and
the mismatch between gravitating masses derived from optical and X-ray
data can be used, for example, to measure the contribution of
non-thermal particles to the gas pressure. Such non-thermal particles
are directly observed as bubbles of relativistic plasma
(e.g., Boehringer et al. 1993). They can also be present in the ICM
due to mixing of thermal and relativistic plasma, or generated by 
shocks propagating through the ICM. Below we use the newest
Chandra data to compare optical and X-ray data in two giant elliptical
galaxies. The $\sim0''.5$ angular resolution of Chandra has
brought the X-ray data on a par with optical data and has made it
possible to extend the comparison between the kinematics of stars and
hot gas from spatial scales of arcseconds up to $\sim 10'$ as is done
below for two giant elliptical galaxies: M87 and NGC1399.

The fundamental assumption that underlies this paper is that the mass
profiles derived from optical observations are correct, so that
deviations between the optical and X-ray profiles are due to effects
on the gas such as non-thermal pressure. While measuring
mass profiles from stellar-kinematic data involves only gravity and
dynamics, uncertainties may remain, mainly because of the
degeneracy between radial variations in mass-to-light ratio and radial
variations in the anisotropy parameter of the velocity ellipsoid
(Binney \& Mamon 1982). Breaking this degeneracy is possible when the
kinematic data provide sufficient information on higher-order moments
of the line-of-sight velocity distribution (LOSVD) and this is used to
constrain the distribution of stellar orbits (Gerhard 1993, Merritt
1993). Uncertainties resulting from modeling good quality absorption
line spectra are $\sim 10\%$ in circular velocity over the range of radii covered (e.g.,
Kronawitter et al. 2000; Thomas et al. 2007, and references
therein). Additional reasons to believe that the optical mass profiles
that we use in this paper are reliable are: (i) the profiles from M87
and NGC1399 are derived from quite different methods---
based on stellar absorption line spectra but including also discrete
tracer particles (globular clusters) in the former case---yet yield
consistent results for the level of non-thermal pressure; (ii) for
both galaxies, the mass distributions have been studied in at least
two independent studies that yield consistent results; (iii) most of
the concerns about the reliability of optical mass profiles arise from
studies with limited observational resolution (e.g., determinations of
black-hole masses), limited radial coverage of data, or
restrictive modeling techniques (e.g., the assumption of isotropic
velocity distributions), neither of which is a concern in the studies
used here. 

\section{Selection of M87 and NGC1399}

The choice of M87 and NGC1399 is motivated by the following
considerations. We want the objects (i) to be nearly round (E0-E1)
elliptical galaxies with existing high quality optical data on stellar
kinematics, (ii) to have X-ray emission dominated by diffuse gas, with
temperatures not  much higher than the velocity dispersion of the
galaxy, (iii) to have high-quality Chandra data; (iv) to exhibit at
most small or moderate deviations from spherical symmetry in the X-ray
images of the hot gas. There are few objects that satisfy all these
criteria. For example, condition (i) requires that the object be
nearby, while (ii) requires that we study massive early-type galaxies
at the center of a group or a cluster with a cool core.  Choosing a
massive galaxy is necessary to avoid contamination of the X-ray
emission by unresolved point sources. For instance in NGC3379 (the
dominant elliptical galaxy in the poor Leo group), the X-ray emission
remaining after removing bright low mass X-ray binaries has a large
(perhaps dominant) contribution from the unresolved stellar population
(CVs, coronally active stars). This conclusion follows from the
comparison of the unresolved X-ray luminosity per unit K-band
optical luminosity of NGC3379 with that for the gas poor bulge of M32
(Revnivtsev et al., 2007, 2008; see also David et
al. 2006). Application of the hydrostatic equilibrium equation to such
systems would lead to the conclusion that the gas has been heated to
a temperature that cannot be bound to the galaxy, and is
out-flowing (Pellegrini \& Ciotti 2006).

Both M87 and NGC1399 are at least several times more massive than
NGC3379.  They reside in the gas-rich cool cores of nearby clusters
(Virgo and Fornax respectively), and both objects have excellent
optical data and long Chandra observations (see
Fig.\ref{fig:image_m87},\ref{fig:image_ngc1399}). Thus, although we
plan to apply our technique to additional galaxies in the future, M87
and NGC1399 satisfy all criteria enumerated above and are the
best targets for our pilot project.

We assume distances of 16 and 19.8 Mpc for M87 and NGC 1399
respectively.

\section{Observations and data preparation}

Our
\emph{Chandra} data analysis is adopted from the reduction procedure
described in \citep[V05 hereafter]{2005ApJ...628..655V}. This includes
filtering of high background periods and application of the latest
calibration corrections to the detected X-ray photons, and
determination of the background intensity in each observation.

\defcitealias{2005ApJ...628..655V}{V05}

The quiescent \emph{Chandra} background is dominated by
the cosmic X-ray background and charged particle events. The
latter component can be subtracted 
exquisitely accurately \citep[with a $\lesssim 2\%$ scatter,
see][]{2006ApJ...645...95H} if one removes the secular trends in
background intensity using the hard X-ray band data.  The cosmic
X-ray background component is modeled
adequately by using the ``blank-sky'' background datasets\footnote{See
http://cxc.harvard.edu/contrib/maxim} which include both the
particle-induced and unresolved sky components.

The spectral calibration includes removal of the time-dependent gain
variations of the ACIS CCDs, and also the most recent corrections for
all positional-dependent variations in the ACIS response and effective
area. For the discussion of associated uncertainties, see
\citetalias{2005ApJ...628..655V}. They are negligible for our purposes.

\subsection{M87 data}
For the analysis we used Chandra observations (OBSIDs 5826, 5827,
5828, 6186, 7210, 7211, and 7212) taken at a variety of instrument
roll angles from February to November 2005 using the ACIS-I detector
(CCDs I0-I3) in Very Faint (VF) mode to minimize the background (see
Forman et al. 2007 for details).  

We reprocessed all observations applying the latest CTI and
time-dependent gain calibrations.  We performed the usual filtering
by grade, excluded bad/hot pixels and columns, removed cosmic ray
`afterglows', and applied the VF mode filtering. We also reprocessed
two Faint mode ACIS-S OBSIDs (3717 and 2707) and treated the front
(S2) and back (S3) illuminated CCDs independently.  We compared the
images from the Very Faint and Faint mode observations to verify that
no artifacts were introduced near the bright jet by the use of the
large $5\times5$ pixel event regions in VF mode. We examined the data
for background flaring and found moderate flaring in OBSIDs 3717 and
2707 (the back illuminated CCDs only) for which we excluded
approximately half the duration.  A typical effective exposure time is
$\sim500$ ksec. The background files (see Markevitch 2001 for details)
were processed in exactly the same manner as the observations.

During 41~ms of the 3.2~s nominal exposure readout of the ACIS
CCDs, the chips are exposed to the sky.  This results in a small
contribution of the source flux, 1.3\%, being uniformly re-distributed
along the readout direction.  This `readout artifact' is most clearly
visible when there is a bright point source in the FOV which produces
a characteristic readout streak. However, for any emission where a
bright region contaminates one of lower surface brightness, this
effect must be taken into account.  The readout artifact can be
accurately subtracted using the technique described by Markevitch et
al. (2000).  A new event file is generated using the original data by
randomizing the CHIPY-coordinate and all sky coordinates and
energies are recalculated as if it were a normal observation. The new
event file is renormalized by the ratio of the readout time compared
to the integration time ($41\,{\rm ms}/3.2\,{\rm s}=0.013$) and then
treated exactly as another component of the background.  The
blank-field background is also renormalized by reducing its
integration time by 1.3\% to account for this additional subtraction.
The M87 nucleus and the brightest jet knot were piled-up in most of
the observations and the corresponding readout streaks were completely
excluded from the analysis.

As the last step of data preparation, an additional column was 
added to each event list. For a given event in the list this column
$\eta$ contains the ratio of the effective area for a photon with
a given energy at a given position $A(E,x_d,y_d)$ to a 
predefined function of energy $A_0(E)$:
\begin{eqnarray} 
\eta=A(E,x_d,y_d)/A_0(E),
\label{eqn:eff}
\end{eqnarray}
where $A(E,x_d,y_d)$ includes mirror and detector efficiencies
(including non-uniformity of the detector quantum efficiency and the
time and spatially dependent contamination on the optical blocking
filter).  Our data set  largely consists of ACIS-I data and a
natural choice is to set $A_0(E)$ to the ACIS-I on-axis effective
area. This makes $\eta$ equal to the vignetting of the mirrors,
modified by the (energy dependent) variations of quantum efficiency
across the detector. The same procedure was repeated with the
background data. Finally, for each event list, we make an exposure map
that accounts for all position dependent, but energy independent,
efficiency variations across the focal plane (e.g., overall chip
geometry, dead pixels or rows, variation of telescope pointing direction).

\begin{figure*} 
\plottwo{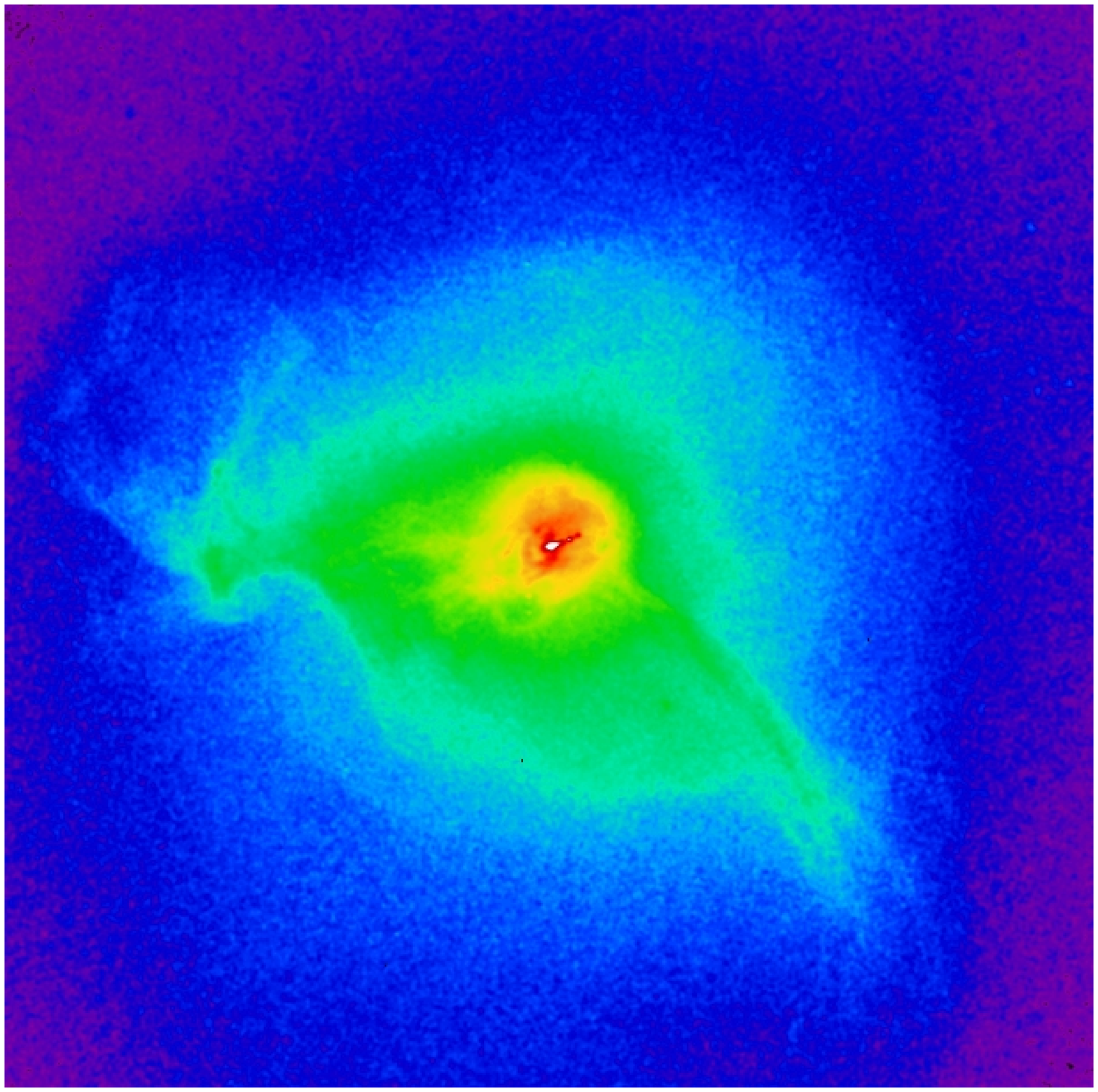}{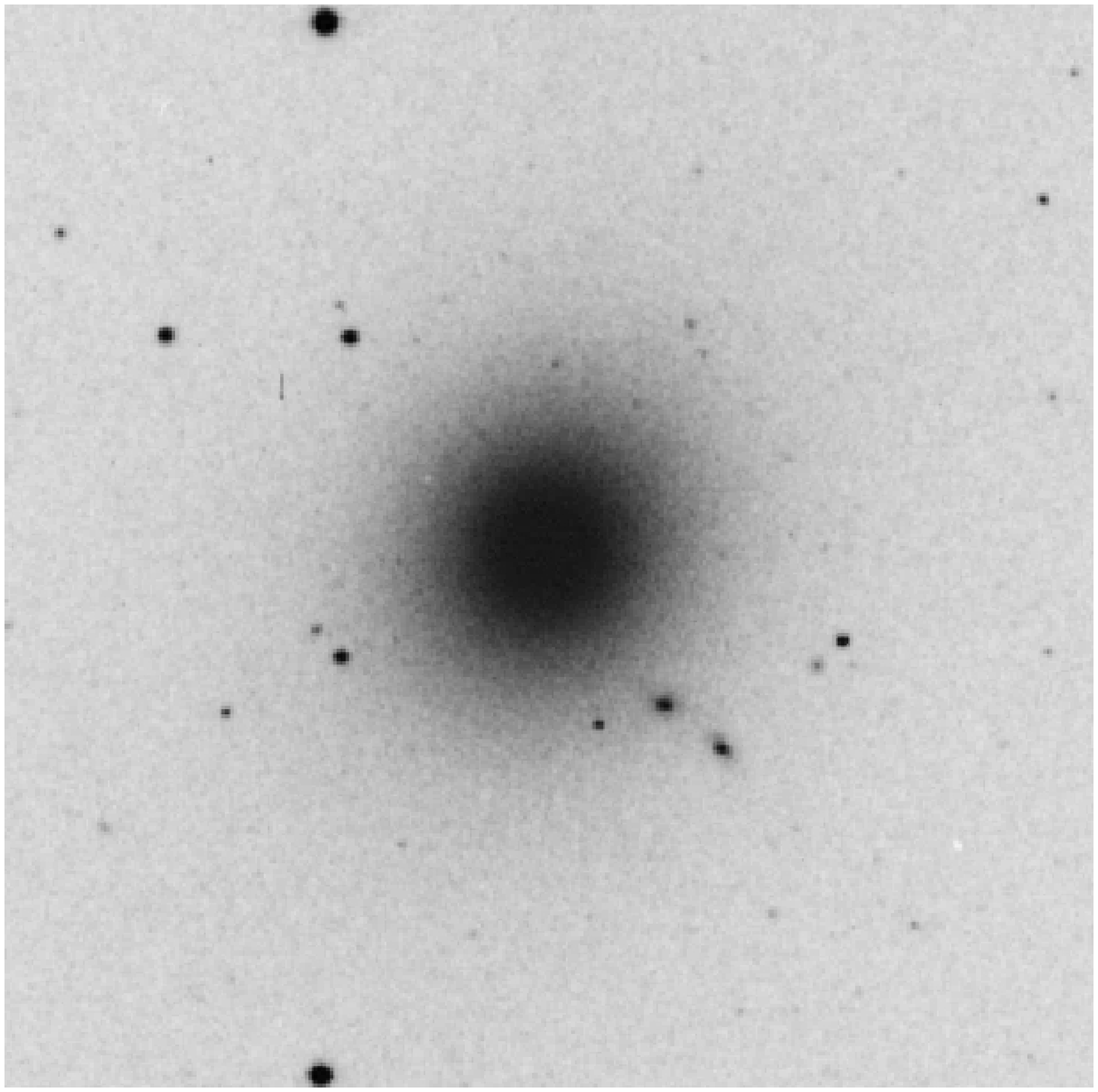}
\caption{Central $10' \times 10'$ ($10'=47$ kpc) region of the Chandra
0.6--2 keV (left) and DSS optical image (right) of M87. The galaxy
appearance is very regular in the optical band, while the X-ray image
is moderately disturbed. In these and other images throughout
the paper, north is up and  east is to the left.
\label{fig:image_m87} } 
\end{figure*}

\begin{figure*}
\plottwo{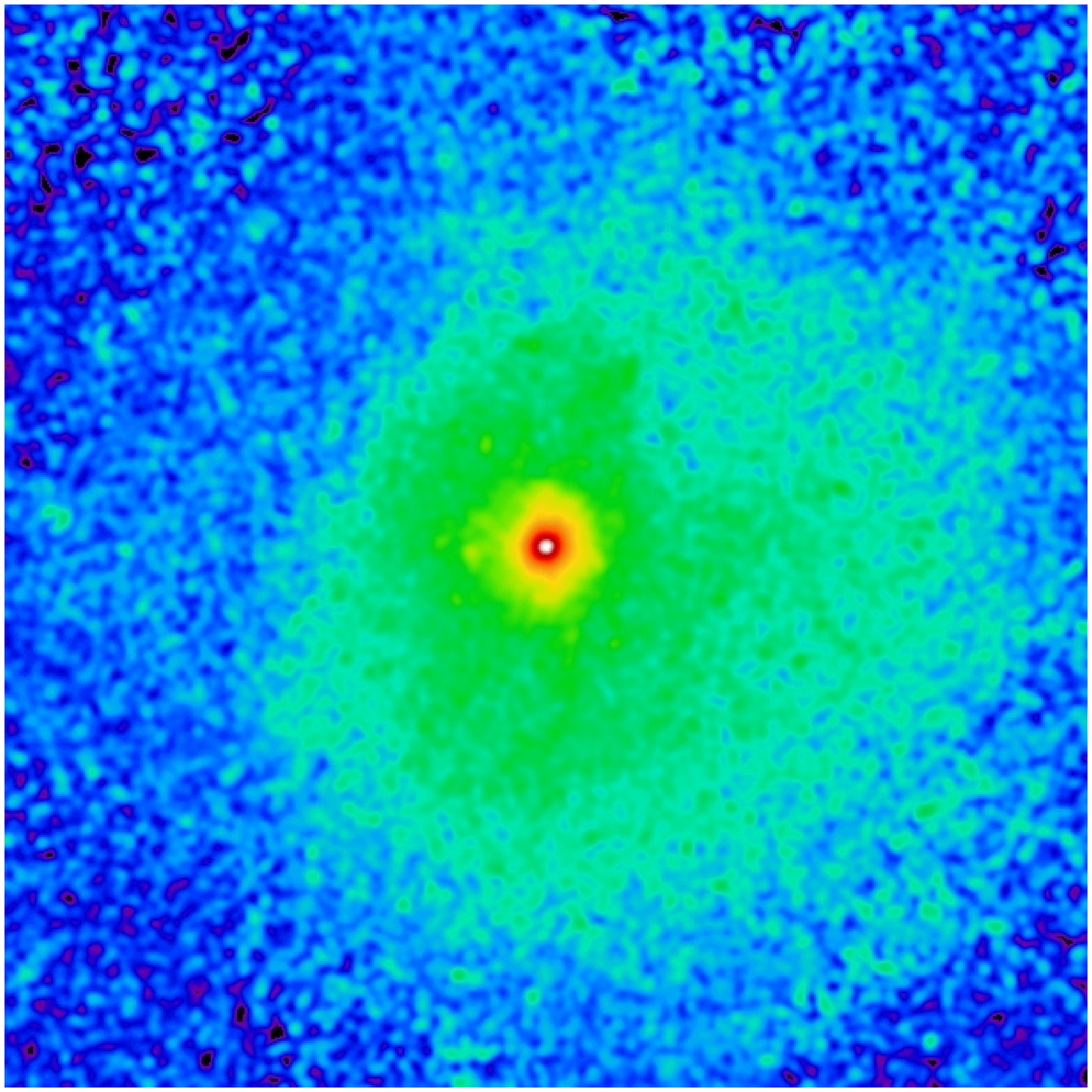}{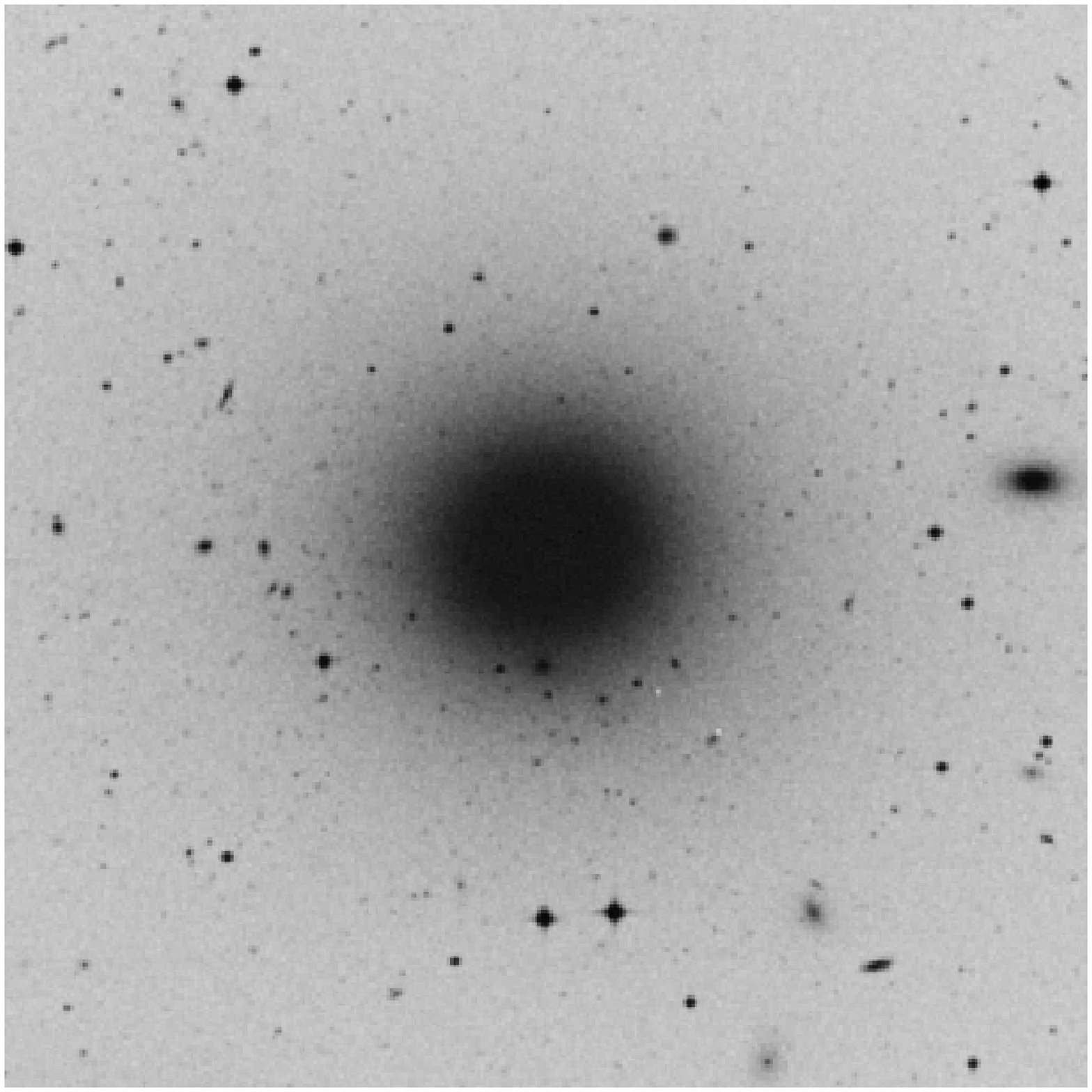}
\caption{Central $10' \times 10'$ ($10'=58$ kpc) region of the Chandra 0.6--2 keV
  (left) and DSS optical image (right) of NGC1399. As for M87, the
  optical galaxy is very regular, while the X-ray image shows
  only moderate deviations from spherical symmetry and appears considerably
  less disturbed than M87.
\label{fig:image_ngc1399}
}
\end{figure*}

\subsection{NGC 1399 data}
For NGC 1399 we used Chandra data of OBSIDs 319, 4172 and 4174, having
a total useful exposure of $\sim$120 ksec. The data were cleaned and
processed using the same procedures as for the M87 data.

\section{Deprojection}
\label{sec:deprojection}
We use the observations from M87 to describe our method in detail and
to explore biases arising from multi-temperature gas components.  Such
components are clearly visible in M87 (see Fig.~\ref{fig:image_m87})
and are less prominent in NGC 1399 (see
Fig.~\ref{fig:image_ngc1399}).  Hence, M87 provides the most stringent
test of our method.

\subsection{Broad band}
\begin{figure}
\plotone{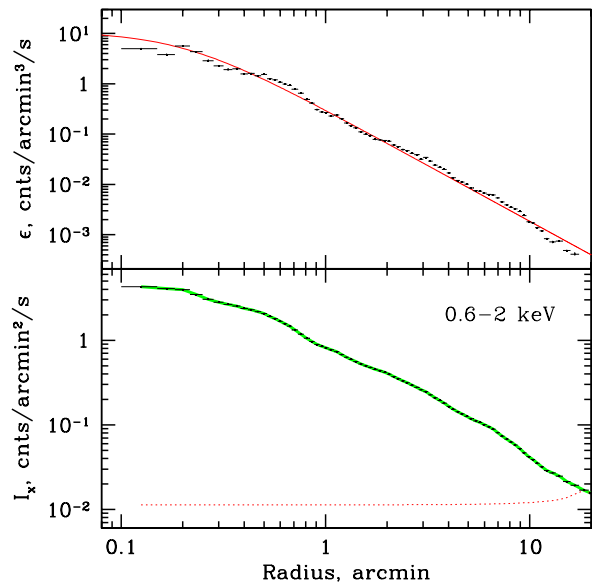}
\caption{M87 surface brightness in the 0.6--2 keV band (lower panel) and 
the emissivity profile in the same band (upper panel). The solid line in 
the upper panel shows the function $\displaystyle 
(1+(r/r_c)2)^{-3\beta}$, with
$\beta=0.37$ and $r_c=0.2'$. Clearly, the $\beta$-model provides a
reasonable, but not perfect, description of the gas emissivity in the
0.6--2 keV band.
The thick green line in the lower panel shows the
projection of the emissivity profile for comparison with the observed 
surface
brightness. It goes through the data points, showing that the
deprojection solution is exact. The nearly
horizontal dotted line is the contribution to the surface brightness in each
annulus from the outer gas layers (outside $19'$), where the gas 
emissivity is
assumed to decline as a power law of radius.
\label{fig:sb}
}
\end{figure}

First, using the event lists for all 9 observations, we build an
image in the 0.6-2 keV band, centered at the M87 nucleus, $40'$ on a side
and with $1''$ pixels. To build this image, the image pixels were initialized
to zero, then for each event in the event list with energy in this range,
we added the value $1/\eta$ to the corresponding image pixel\footnote{
Since $\eta$ is essentially the vignetting factor (relative to the
on-axis position), populating images with $1/\eta$ instead of $1$
for each event is equivalent to applying the vignetting correction 
during image construction instead of accounting for this factor in the
exposure map. The logic behind this approach is further discussed in
Section \ref{sec:dspec}.}. The
same procedure was applied to the background event lists. An exposure
map was generated combining exposure maps for individual event lists.
A background-corrected surface brightness profile $S_{obs}(i)$ 
calculated from the `$1/\eta$' image in a set of annuli centered at
the M87 nucleus is shown in the bottom panel of Fig.\ \ref{fig:sb}. The
surface brightness is normalized to counts per arcmin$^{2}$ per
second.

In our deprojection analysis, we follow the approach used in Churazov et
al. (2003) (see Appendix \ref{sec:deproj} for additional discussion). We assume
spherical symmetry, but make no specific assumption about the form of the
underlying gravitational potential. For a given surface brightness profile in
$n_a$ annuli, we choose a set of $n_s$ ($n_s\le n_a$) spherical shells
with the inner radii
$r(i),~i=1,\ldots,n_s$. The gas emissivity $\mathcal{E}$ is assumed to be
uniform inside each shell, except for the outermost shell, where the
gas emissivity  is assumed to decline as a power law of radius: $\mathcal{E}=
\mathcal{E}_{out} r^{-6\beta_{out}}$, where $\beta_{out}$ is a parameter. In
our analysis $r(n_s)$ was set to $19'$ and $\beta_{out}=1/3$. In practice, we are mostly working with the data inside a $10'$
circle and the precise values of $\beta_{out}$ and the outer radius do not
affect the results.

The expected surface brightness can then be written as a projection of
emissivities in each of the $n_s-1$ shells plus a contribution from the outer
layers ($r>r(n_s)$): 
\begin{eqnarray}
S(j)=\sum_{i=1}^{n_s-1} P(i,j)\mathcal{E}(i)+P_{out}(j)\mathcal{E}_{out},
\end{eqnarray}
where $\mathcal{E}(i)$ is the emissivity of a given shell,
$\mathcal{E}_{out}$ is the emissivity of the outer layers of the gas at
$r=1$, and $P(i,j)$ and $P_{out}(j)$ are the projection matrix/vector from
our set of shells into our set of annuli. A simple analytical
expression for $P(i,j)$, which is a function of the geometry only,
is given by McLaughlin (1999).

Deprojection can then be reduced to a simple least squares problem -- what
set of emissivities in our set of shells (together with the
emissivity normalization $\mathcal{E}_{out}$  for the outer layers)
provides the best description of the observed surface brightness:
\begin{eqnarray}
\chi^2=\sum_{j}\left [S(j) - S_{obs}(j) \right ]^2 /\sigma(j)^2 =
min
\label{eq:chi2}
\end{eqnarray}
where $\sigma(j)$ is the error associated with the surface brightness
in annulus  $j$. In our analysis, we use modified errors
$\sigma(j)^2=\sigma_{stat}(j)^2+\delta^2\times S^2_{obs}(j)$, where
$\sigma_{stat}(j)$ is the statistical error associated with the
Poisson noise in the observed image and background and $\delta\sim
0.1$. This was done to avoid the situation where a few annuli with the
best statistics completely dominate the $\chi^2$. In practice for the M87
dataset, the choice of $\delta$ does not affect the results. As usual
the differentiation of this relation with respect to $\mathcal{E}(i)$
and $\mathcal{E}_{out}$ yields a system of linear equations
$\tilde{P}\mathcal{E}=\tilde{S}$, which can
easily be solved. Here $\tilde{P}$ is the square $n_s\times n_s$
matrix and each of the $n_s$ elements of $\tilde{S}$ vector are the
linear combinations of the original observed values of $S_{obs}$. 
In the cases considered below, the solution of 
equation (\ref{eq:chi2}) is unique, once the parameter $\beta_{out}$ is
fixed. The properties of the inverse matrix $\tilde{P}^{-1}$ (in
particular an error enhancement when the ratio of maximal to minimal
eigenvalues of $\tilde{P}$ is large) can easily be controlled by making the
spherical shells broader. A practical recipe for the choice of
shells/annuli radii will be described elsewhere.

The emissivity of each shell can then be evaluated as an explicit linear
combination of the observed quantities: 
\begin{eqnarray}
\mathcal{E}(i), \mathcal{E}_{out}=\sum_{j=1,n_s} \tilde{P}^{-1}(j,i) \tilde{S}(j)=\sum_{j=1,n_a} D(j,i) S_{obs}(j),
\label{eq:dprj}
\end{eqnarray}
where $D$ is the final $n_a\times n_s$ deprojection matrix
relating the observed surface brightness in rings and the emissivities
of spherical shells. 
Since the whole procedure is linear, the errors in the observed
quantities can be propagated straightforwardly. Shown in
Fig. \ref{fig:sb} (upper panel) is the set of gas emissivities in the
0.6-2 keV band derived from the surface brightness profile. The
forward convolution of this set of emissivities with the projected
matrix $P(i,j)$ (green line in the lower panel of Fig. \ref{fig:sb})
exactly reproduces the observed surface brightness profile. 

\subsection{Deprojected spectra}
\label{sec:dspec}
Note that with our definition of $\eta$, the projection matrix $P(i,j)$
does not depend on energy. Indeed all energy/position dependent
factors are already corrected for when we make the $1/\eta$ image
instead of a usual image constructed from raw counts without any
weights. We do lose some of the sensitivity with this $1/\eta$
procedure, but very little as long as the variations of $1/\eta$ across
the region used for spectrum extraction are not large. Note that
since the exposure time is dominated by that from the ACIS-I chips, 
the factor $\eta$ is a smooth function of energy and our
$1/\eta$ technique does not introduce spurious spectral
features.

\begin{figure}
\plotone{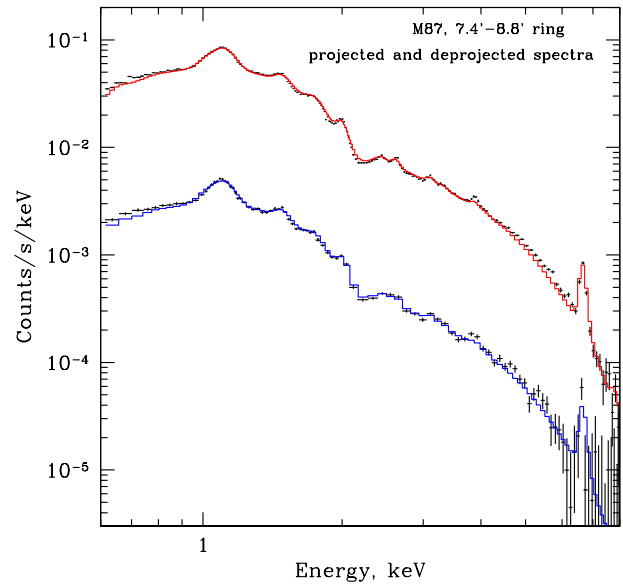}
\caption{Projected (top) and deprojected (bottom) spectra 
for the $7'.4-8'.8$
annulus/shell in M87 (black crosses) together with the best 
fitting single-temperature
APEC model (red and blue histograms). The best fitting 
temperatures are 2.57 and 2.4 keV for
projected and deprojected spectra respectively.  Both spectra are
reasonably well described by a single temperature APEC model.
\label{fig:spec}
}
\end{figure}

Since $P(i,j)$ does not depend on energy, the deprojection of a
surface brightness profile in any energy band would look like equation
(\ref{eq:dprj}) with exactly the same $D(j,i)$.   Thus we can
accumulate a set of `$1/\eta$' spectra (corrected for background and
readout) for each of the $n_a$ annuli, and apply equation\ 
(\ref{eq:dprj}) to determine the emissivities of each shell in each of
the ACIS energy channels.   We
note here that a similar technique of vignetting correction at the
level of individual events is often used in the analysis of X-ray data
(e.g. for XMM-Newton). For M87 Chandra observations we  explicitly verified
that  fitting of `$1/\eta$' and 'regular' spectra of the same regions
yields fully consistent results.

 The procedure described above differs slightly from the often
used onion peeling algorithm or  \verb#projct# in XSPEC (Arnaud 1996).
A short note on the different deprojection techniques is given in the
Appendix~\ref{sec:deproj}. The bottom line is that all techniques should
yield consistent results if underlying assumptions are correct. Our
procedure is computationally fast and is especially useful when a large
number of spherical shells are analyzed.

\begin{figure}
\plotone{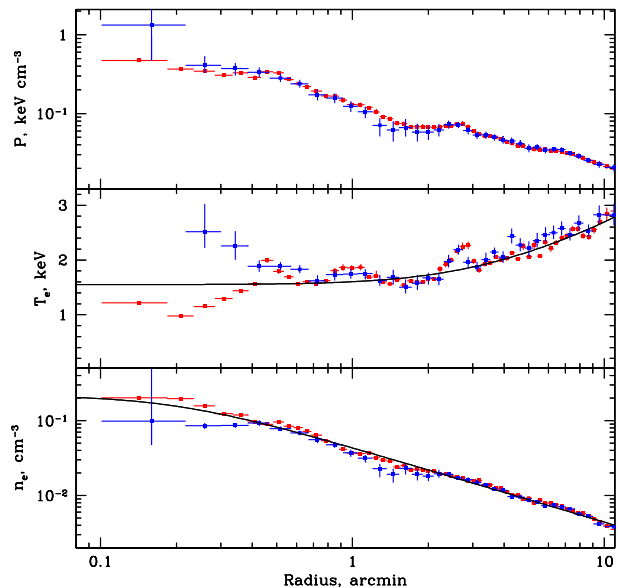}
\caption{Radial profiles of deprojected gas parameters in M87. The
parameters were obtained using a single temperature APEC model (with
fixed low energy absorption and redshift) fit to the deprojected spectra
for a set of spherical shells centered on M87.  The red and blue
points correspond to the best-fitting parameters for the 0.6--9 keV and 
2.5--9
keV energy bands respectively. Simple analytical approximations for
the density and temperature are shown with the solid lines (eqs. \ref{eq:ne}
and \ref{eq:te}). The
uppermost panel shows the gas pressure evaluated as $P=nT=1.91n_e T_e$,
where $n_e$ and $T_e$ are the best-fitting electron density and
temperature. Outside the central $0'.4$ the best fit parameters
for the two bands agree well, suggesting that the impact of
multi-temperature plasma on the $n_e$ and $T_e$ is not strong.
\label{fig:dspec}
}
\end{figure}

The resulting spectra were fitted in XSPEC V12 (Arnaud 1996) with a
single-temperature optically thin plasma emission model as implemented
in the APEC code (Smith et al. 2001). The low energy photoabsorption
was fixed at the Galactic value $N_H=2.54 \times 10^{20}~{\rm
cm}^{-2}$ (Dickey \& Lockman 1990) and the redshift was fixed at the
optically determined redshift of M87: $z=0.00436$ (e.g., Smith et
al. 2000). The gas temperature, heavy metal abundance and
normalization were free parameters of the model. A sample spectrum
in the $7'.4-8'.8$ annulus and the deprojected spectrum in the
spherical shell of the same size are shown in Fig.~\ref{fig:spec}
together with the best fitting single-temperature APEC model. The best
fitting temperatures are 2.57 and 2.4 keV for projected and deprojected
spectra respectively.  The deprojected spectra were generated in
narrow (14.6 eV) energy bins and used in XSPEC without any further
grouping. To avoid a bias towards lower temperatures, which is
present when the number of counts per spectral bin at high energies
becomes less than $\sim 10-20$, the statistical errors were evaluated
from the smoothed spectra, while the fitting is applied to raw
(unsmoothed) spectra (see Churazov et al. 1996 for
details). This approach suppresses the bias associated with the low
number of counts per spectral bin provided that total number of counts
in the spectrum is large. We did verify earlier for ASCA (Churazov et
al. 1996) and XMM-Newton data
and now for Chandra observations of M87 that this approach yields fully
consistent results to an often used technique of grouping spectral
channels in bins with at least 10-20 counts (assuming that the correct
spectral model is used). Note that we also verified that allowing
the Fe abundance to vary separately from the $\alpha$-elements altered
the fits by not more than the statistical errors.

The electron density was derived from the normalization of the
spectra, fixing the proton to electron ratio to $0.83$ and taking into
account the distance to M87. The resulting spectral parameters are plotted
in Fig.\ \ref{fig:dspec} as a function of distance from the center of
M87. Red and blue points correspond to the best-fitting model parameters in
the 0.6--9 and 2.5--9 keV bands respectively. This choice of two energy
bands for spectral fitting is motivated on the following grounds: the
broader band maximizes the statistical significance of the results, while
the harder band provides a verification of the magnitude of biases
which might arise when a one-temperature model is applied to a
spectrum with multiple temperature components (e.g., Buote 2000).

\begin{figure}
\centering \leavevmode
\plotone{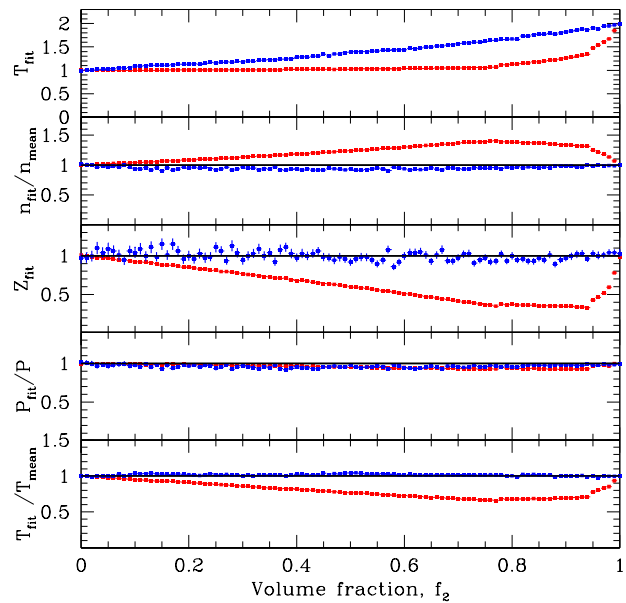}
\caption{`Biases' in a two-temperature plasma versus volume fraction
of the hotter component. Fake Chandra spectra representing a
mixture of $kT_1=1$ and $kT_2=2$ keV plasmas in pressure equilibrium
were fitted with a one-temperature model in the 0.6--9 and 2.5--9 keV
bands (red and blue points respectively). The five panels from top
to bottom show i) the best-fitting gas temperature, ii) the ratio of
best-fitting gas density to the true mean density, iii) gas
metallicity, iv) gas pressure $P_{fit}\equiv n_{fit}T_{fit}$, v) the ratio of the measured value 
$P_{fit}/n_{fit}\equiv T_{fit}$ to the true value
$P/n_{mean}\equiv T_{mean}$; here $n_{mean}$ is the volume averaged plasma
density (eq.\ \ref{eq:nmean}). The last quantity characterizes the bias in
measuring the gravitating mass or the potential from the hydrostatic equilibrium
equation.
\label{fig:bias}
}
\end{figure}

To assess the magnitude of possible biases when fitting the multi-temperature
plasma in the Chandra band with a one-temperature model, we generated a
sequence of fake spectra representing a mixture of $kT_1=1$ and $kT_2=2$ keV
plasmas in pressure equilibrium, with a varying volume fraction $f_2$ of the
harder component.  These two temperatures are characteristic for
multi-temperature plasma in the inner parts of M87 and NGC1399 (Buote
2002). The abundance of heavy elements was set to solar for both
components, and $N_H$ and redshift were fixed to the M87 values.   The on-axis
ACIS-I response was used. We then applied to the simulated spectra the same
one-temperature model as we did for the real M87 data. The resulting spectral
parameters are shown in Fig.\ \ref{fig:bias}: temperature, density,
metallicity, pressure $P_{fit}\equiv n_{fit}T_{fit}$, and the ratio of the measured value 
$P_{fit}/n_{fit}\equiv T_{fit}$ to the true value
$P/n_{mean}\equiv T_{mean}$; 
\begin{equation}
 n_{mean}=n_{1}\times (1-f_2)+n_{2}\times f_2
\label{eq:nmean}
\end{equation}
is the volume averaged plasma density and $n_{1}$ and $n_{2}$ are the
plasma densities  of the cooler and hotter phases respectively
(evaluated as $1.91$ times the respective electron densities $n_e$). The quantity
$T_{fit}/T_{mean}-1$ characterizes the bias in measuring gravitating mass or
potential from the hydrostatic equilibrium equation. The red and blue
points correspond to the best-fitting parameters derived from the 0.6--9 and
2.5--9 keV bands respectively. For the 0.6--9 keV band, even a very small
volume fraction of the cooler component causes the best-fitting temperature
and metallicity to drop sharply, while the density rises (red points in Fig.\
\ref{fig:bias}). For the 2.5--9 keV band, the density is, on the contrary,
very close to the mean value. This conclusion is of course sensitive to the
particular choice of $T_1$ and $T_2$ and to the effective area of the
instrument. In general, one expects the best-fitting temperature in the broad
band to be more biased towards lower values compared to the hard band. It is
possible therefore to use the differences in the best-fitting values obtained
in two energy bands as an indicator of strong departures from a
single-temperature plasma. From Fig.\ \ref{fig:dspec} it is clear that the
best-fitting values in the 0.6--9 and 2.5--9 keV bands differ strongly inside
the inner $0'.4$, while outside this radius, the results agree well. Of course
even a reasonable agreement between the spectral results in two bands does not
guarantee that the spectrum is fully described by a one-temperature
model. This agreement nevertheless suggests that outside $0'.4$ our
spectral results are not strongly biased. On the other hand, the region inside
$0'.4$ is characterized by very strong departures from spherical symmetry and
by the presence of cool and hot gas structures. Given the magnitude of
the expected biases the results for this region should be taken with caution.

\begin{figure}
\plotone{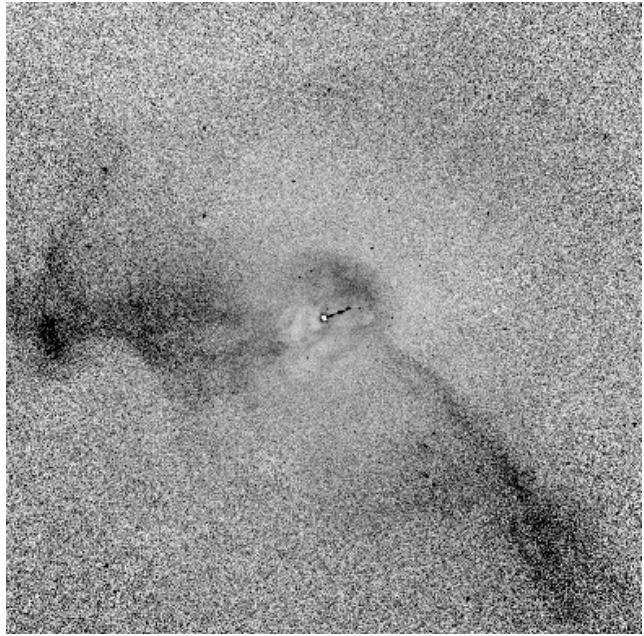}
\caption{ Central $10' \times 10'$ part of the 0.6--2 keV image of M87,
divided by the axisymmetric model with surface brightness $I(R)
\propto (1+\left(R/r_c\right)2)^{1/2-3\beta}$ (here $R$ is the
distance from the M87 nucleus, $\beta=0.37$ and $r_c=0.2'$) to show the
structure in the surface brightness more clearly.
\label{fig:m87dbimage}
}
\end{figure}

From Chandra and XMM-Newton observations we know that the gas
temperature in the X-ray bright features co-spatial with the extended
radio bright `arms' is lower than that of the typical intra-cluster
medium (ICM) temperature at the same distance from the center (e.g.,
Belsole et al. 2001; Matsushita et al. 2002; Forman et
al. 2005). These regions are clearly seen in Fig.\
\ref{fig:m87dbimage}, where the 0.6--2 keV image of M87 is
divided by an axisymmetric $\beta$-model, as extended structures to the east and
the south-west of the nucleus. To test whether these structures
dramatically affect our final result (e.g., the mass/potential
profile, discussed below) we also fit the data excluding those regions
containing the brightest parts of the arms. We use the spectral fits
to this `trimmed' data-set in the next section to verify the
robustness of our results.

Analytical approximations to the best-fitting values of $n_e(r)$ and $kT(r)$
over the range of radii from $\sim 0.5'$ to $\sim 10'$ are given below:
\begin{eqnarray}
n_e=0.22~\left[1+\left(\frac{r}{r_c}\right)^2\right]^{-\frac{3}{2} \beta} {\rm ~~cm^{-3}},
\label{eq:ne}
\end{eqnarray}
where $\beta=0.33$ and $r_c=0.2'$ (or 0.93 kpc).

\begin{eqnarray}
kT=1.55\times\left[1+\left(\frac{r}{2.2'}\right)^2\right]^{0.18} {\rm ~~keV}.
\label{eq:te}
\end{eqnarray}

These approximations are shown in Fig.\ \ref{fig:dspec} by the
black solid lines.

\section{Potential}
\label{sec:pot}
We first derive the gravitational potential for M87, including a
detailed discussion of the method which we then apply to the data for
NGC 1399.

\subsection{M87}
\label{sec:m87}
We assume that the gaseous atmosphere of M87 is in hydrostatic
equilibrium:
\begin{eqnarray}
\frac{1}{\rho}~\frac{dP}{dr}=-\frac{d\varphi}{dr}
\label{eq:hyd}
\end{eqnarray}
where $\rho$ is the gas density, $P=P(r)$ is the pressure and $\varphi$ is the
gravitational potential.  Thus
\begin{eqnarray}
\varphi=-\int \frac{1}{\rho}~\frac{dP}{dr} dr
\label{eq:pot0}
\end{eqnarray}
If the pressure is solely due to the thermal gas pressure, then
$P=nkT$ and $\rho=\mu m_p n$, where $\mu$ is the mean atomic weight of
the gas particles, $m_p$ is the proton mass, and $n=n(r)$ is the total
particle density.  Then (assuming that $\mu$ does not vary with
radius) the potential $\varphi$ is:
\begin{eqnarray}
\varphi=-\frac{k}{\mu
m_p}\left[ \int{ T\frac{d\ln n}{dr} dr} +T \right]+C 
\label{eq:pot}
\end{eqnarray}
where $C$ is an arbitrary constant. Below we will use the value of the
potential relative to a given reference radius $r_{ref}=5'.5$ (25.6 kpc), i.e.,
$\varphi(5'.5)=0$.  Note that eq.~(\ref{eq:pot}) depends only on the measured
quantities $n(r)$ and $T(r)$. Furthermore, eq. (\ref{eq:pot}) does not depend
explicitly on radius $r$ (only through $n(r)$ and $T(r)$) or on the absolute
normalization of the gas density. This in particular means that the assumed
distance to M87 does not affect the calculated potential as long as it is
expressed as a function of angular distance from the nucleus.  Thus $\varphi$
can be easily evaluated using the densities and temperatures derived from the
deprojected X-ray spectra. Unlike the expression for gravitating mass, which
explicitly depends on the spatial derivatives of the observed quantities, the
change of the potential $\Delta \varphi$ between two radii $r_1$ and $r_2$
depends primarily on the gas densities at two radii and on the mean gas
temperature. For example, in the case of isothermal gas, $\Delta
\varphi \propto T \ln \left[n(r_2)/n(r_1)\right)$ which is of
course a direct consequence of the Boltzmann distribution.  This makes the
calculation of $\Delta \varphi$  very simple and
robust. 

Below we will denote the potential derived from X-ray data (by
means of eq.~\ref{eq:pot}) as $\varphi_X$ to distinguish it from the true
potential $\varphi$ or from the potential $\varphi_{opt}$ derived from
optical data.

Note that the derived spectral parameters in each bin are not independent
random variables since they come from the deprojection analysis, e.g., the
emissivities in the spherical shells are various linear combinations of the
same set of original independent data points (fluxes in the concentric annuli
around the source). For a given spherical shell the errors are evaluated
correctly since the spectrum is calculated as an explicit linear combination
of independent variables.  However, it is expected that the deviations from
the correct values in nearby radial shells are
correlated/anti-correlated. Thus, if we repeat the same observation many times,
the measured values of the emissivity will lie within the estimated
uncertainties in 67\% of cases (apart from possible systematic errors), but
the values in nearby radial bins may change in correlated/anti-correlated ways.

Similarly, the values of $\varphi_{X}$ obtained at different radii are not
statistically independent, both because they are derived from the deprojected
spectral parameters and because the value of $\varphi_{X}$ at each radius
depends on the same set of spectral parameters in various radial bins
according to equation (\ref{eq:pot}). Rigorous error calculations would require
error propagation from the original (independent) spectra obtained in the set
of annuli around the source; these would be difficult to do and the results
would be difficult to present simply. However, in practice we believe this
more rigorous analysis  may be replaced with a straightforward
error calculation. Indeed, for the
case of an isothermal gas $\varphi_X (r) = \frac{k}{\mu m_p}T \ln n(r) +C$, i.e., $\varphi_X
(r)$ depends only on the local (at radius $r$) gas density. Therefore, the error distribution for
$\varphi_X(r)$, at any given radius $r$, can be correctly estimated, as
long as the errors in $n(r)$ are correct (see previous paragraph).
 The same is approximately true for a slowly varying
temperature profile. Below, we assign to each $\varphi_X (r)$ the error
which follows from equation~(\ref{eq:pot}, assuming that the
errors in evaluating the spectral parameters at each radius are independent.

\begin{figure}
\plotone{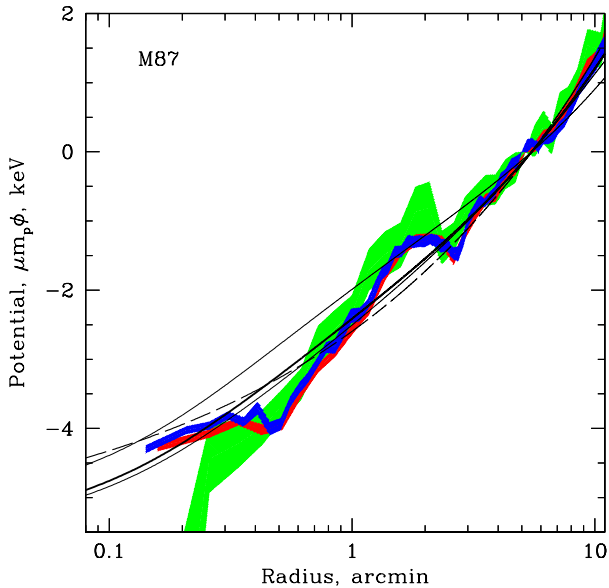}
\caption{Derived potential $\varphi_X(r)$ using the 0.6--9 keV band
(blue), the same excluding `cool arms' (red), the same in the 2.5--9
keV band (green) -- see Table \ref{tab:bands}. Black curves -
potentials $\varphi_{opt}(r)$ derived from the optical data: Wu \&
Tremaine (2006) - dashed line; Romanowsky \& Kochanek (2001), NFW2
model - thick solid line.  The comparison of the thick black and
colored curves suggests that $\varphi_{opt}(r)$ and $\varphi_X(r)$
broadly agree over the range of radii shown in the figure.
 Thin solid lines are the models NFW1
and NFW3 from RK01 which in terms of likelihood/significance are
$\sim$ 0.8 and 0.9 $\sigma$ ``dispreferred'' compared to the best fitting
NFW2 model. The difference between these curves approximately
characterizes the uncertainties in $\varphi_{opt}(r)$ based on the
analysis of RK01. All curves are normalized to zero at $5'.5$ radius. 
\label{fig:pot1}
}
\end{figure}

In Fig.\ \ref{fig:pot1} we show three versions of $\varphi_X$  derived
from the Chandra data for M87. Each of the three shaded `curves' corresponds
to different subsets of data and/or different energy
bands used for spectral fitting, as indicated in Table
\ref{tab:bands}.  The regions that exclude 'arms' are described more fully at
the end of \S4.2. The vertical width of each of the color curves in
Fig.\ \ref{fig:pot1} corresponds to the 67\% statistical uncertainty.

\begin{table}
\centering
\caption{Data used to derive the potential.\label{tab:bands}}
\begin{tabular}{lrr}
\hline
Regions   &  Energy band & Color in Fig.\ \ref{fig:pot1} \\
\hline
0-360$^\circ$ annuli & 0.6--9 keV & blue \\
0-360$^\circ$ annuli, excluding `arms'& 0.6--9 keV & red \\
0-360$^\circ$ annuli, excluding `arms'& 2.5--9 keV & green \\
\hline
\end{tabular}
\end{table}
All three methods lead to estimates of $\varphi_X(r)$ that are consistent with
each other within statistical uncertainties. The potential derived from the
data in the 2.5--9 keV band has
the largest statistical errors, but as discussed above, the spectral
parameters may be less biased in this band, compared to the broader 0.6--9
keV band.  We note here that departures from spherical symmetry are obvious
from Fig.\ \ref{fig:image_m87} and we cannot expect the spectral parameters (or
potential profiles) to be perfectly consistent with each other.

For comparison to the X-ray derived gravitational potentials, we show
in Fig.\ \ref{fig:pot1}, with the black lines, two estimates of the
gravitational potential that are derived from the optical data. The
dashed line shows the potential derived by Wu \& Tremaine (2006, their
\S4.3.2; WT06 below) using
a sample of 161 globular clusters in M87. The mass profile in this
model is a power law function of radius $M(r)=2.3\times 10^{10}~
M_\odot(r/\hbox{kpc})^{1.36}$.  This mass distribution is most
accurately determined near $r=30\,\hbox{kpc}$, but Wu \& Tremaine
estimate that the relative error in $M(r)$ is less than 40\% for
$17\,\hbox{kpc}< r<90\,\hbox{kpc}$ (or $3.7'<r< 19'$). The error in
the potential is substantially smaller, since it is determined by the
integral of the mass distribution, $\phi_{opt}=\int_{r_{ref}}^r
GM(r)dr/r^2$. The potential in the inner few arcminutes is only weakly
constrained since this model does not use any information on stellar
kinematics. The thick solid (black) line in Fig.\ \ref{fig:pot1} is the
best-fitting model to the star and globular-cluster kinematics in M87
from Romanowsky \& Kochanek (2001, see their \S4.2; hereafter, RK01).  This
model (NFW2 in their Table 2) has a mass profile that is the sum of
an NFW profile (Navarro, Frenk \& White, 1996) from a dark halo and a
stellar component, derived from the M87 optical surface brightness
under the assumption of a constant stellar mass-to-light ratio. In
terms of the likelihood, this is the most probable model among those
considered by RK01.  This model agrees reasonably well with the
X-ray derived potential apart from a few clear wiggles in
$\varphi_X(r)$, which we discuss in Section
\ref{sec:shock}.   Two other 
models -- NFW1 and NFW3 (see Table 2 of RK01) are shown in Fig.\
\ref{fig:pot1} with the thin solid lines. In terms of
likelihood/significance these models are $\sim$ 0.8 and 0.9 $\sigma$
``dispreferred'' compared to the best fitting NFW2 model. The
difference between these curves can be used to approximately characterize the
uncertainties in $\varphi_{opt}(r)$ based on the parametric model of RK01.

 We use below the power law model of WT06 and the 
NFW2 model of RK01 as two versions of $\varphi_{opt}(r)$ and assume
that the difference between these two models is comparable to the
deviations of either from these models from true potential of the galaxy.

\subsection{NGC 1399}
\label{sec:ngc1399}
\begin{figure}
\plotone{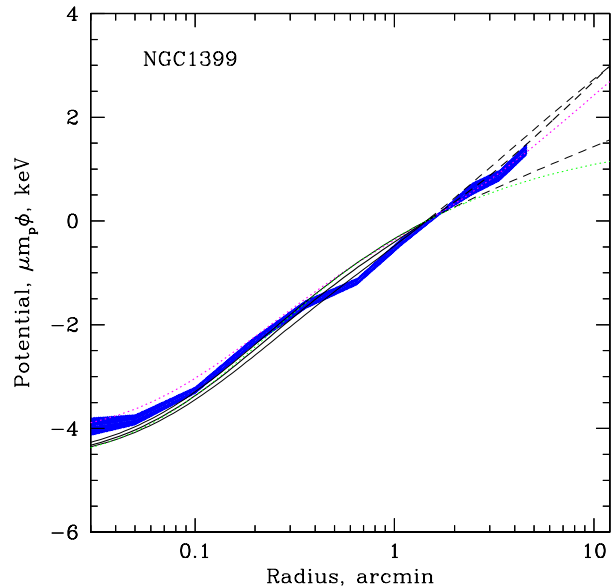}
\caption{Derived potential $\varphi_X(r)$ for NGC 1399 using the 0.6--9 keV
band for 0-360\deg annuli (blue curve). The thick black solid line
shows the potential $\varphi_{opt}(r)$ derived from optical data of
Kronawitter et al. (2000), while thin black lines show $\pm 2\sigma$
limits. Outside the range of radii covered by the optical data ($r<1.62'$),
the extrapolation of the potential is shown by dashed lines.
The green curve is the `stars-only' model from Kronawitter et
al. (2000). The magenta line shows the optical potential
$\varphi_{opt}(r)$ multiplied by 0.9. All curves are normalized to
zero at $1'.5$. The good agreement of $\varphi_{opt}(r)$ and
$\varphi_{X}(r)$ suggests that the fraction of non-thermal pressure is
low.
\label{fig:ngc1399}
}
\end{figure}

The X-ray potential $\varphi_X(r)$ derived from Chandra data for NGC 1399 is
shown in Fig.\ \ref{fig:ngc1399} with the blue curve. The potential is
set to zero at $1'.5$. In X-rays the core
of NGC 1399 appears less disturbed than M87 and the resulting
$\varphi_X(r)$ is much smoother than for M87. For comparison
$\varphi_{opt}(r)$ based on the model of Kronawitter et al. (2000) is
shown with the black solid line,  and their $2\sigma$ confidence
bands as the thin black lines. As in the case of M87, there is good
agreement of $\varphi_X(r)$ and $\varphi_{opt}(r)$ over the region where
X-ray and optical data are available. 
 
The optical potential $\phi_{opt}$ for NGC 1399 was derived from
absorption-line kinematic data, as described in Saglia et al. (2000).
These authors constructed spherical models for the galaxy in a
sequence of gravitational potentials, composed of the potential of the
luminous stars and various optional quasi-isothermal dark halo
potentials. The potential of the stars was based on the deprojected
luminosity distribution and the assumption of constant mass-to-light
ratio. For each of these potentials, the non-parametric distribution
function $f(E,L^2)$,  that gave the best fit to the kinematic
data  (velocity dispersion and symmetric $4^{th}$ moment
$h_4$ of the LOSVD), was found; here $E$ and $L$ denote energy and total angular momentum. The
$\Delta\chi^2$ values of the various models then determined the
boundaries of the confidence region for the halo potential
parameters. Similar models, based on a larger basis of distribution
functions and potentials, were constructed by Kronawitter et
al. (2000). The Kronawitter et al. models are plotted in
Fig.~\ref{fig:ngc1399} and span the 95\% confidence range within their
expanded set of potentials. The true confidence range based on
non-parametric potentials is likely to be somewhat larger; however, it
is not straightforward to determine. An independent investigation by Graham et
al. (1998) yields similar results. 

\section{Effects of Magnetic Fields, Cosmic Rays, and
Micro-turbulence}
\label{sec:eff}
The good agreement between the gravitational potentials derived
independently from the optical and X-ray data, found in the previous
section, suggests that hydrostatic equilibrium is satisfied reasonably
well in M87 and NGC1399. This conclusion contrasts with that of Diehl
\& Statler (2007) who have argued recently that hydrostatic
equilibrium of the X-ray gas in elliptical galaxies is ``the exception
rather than the rule'' and ``X-ray-derived radial mass profiles may be
in error by factors of order unity''.   In particular, NGC1399 is
one of the objects from the Diehl \& Statler (2007) sample that
is identified as being far from hydrostatic equilibrium based on
differences in the isophotal shapes of the stars and the X-ray
gas. We believe that the close agreement of the potential profiles
$\phi_X(r)$ and $\phi_{opt}(r)$ obtained in the previous section
would be an improbable coincidence if hydrostatic equilibrium were
not approximately valid. While neither NGC1399 nor M87 are in
perfect equilibrium, as indicated by visibly disturbed X-ray images
(see e.g. Buote \& Tsai, 1995), the estimate of the magnitude of the
departure from hydrostatic equilibrium by Diehl \& Statler (2007) is
too pessimistic. We will further test this statement in a larger
sample of elliptical galaxies in future work.

We now consider how magnetic fields, cosmic rays and micro-turbulence would
affect the potential $\varphi_X(r)$ derived from the X-ray data. We keep the
hydrostatic equilibrium assumption (eq. \ref{eq:pot0}), but we now allow the
pressure or density derived from X-ray analysis to differ from the true $P$
and $\rho$ entering eq. (\ref{eq:pot0}).

\subsection{Uniform medium}
\label{sec:uniform}
Let us first keep the assumption that the X-ray emitting gas is
uniform within each spherical shell, but assume that this gas provides
only part of the total pressure. The rest of the pressure is provided
by a component that is `invisible' in X-rays (e.g., cosmic rays), which
is also uniformly distributed through the volume of each shell. Here
and below we assume (unless stated otherwise) that the pressure is
isotropic and the atmosphere is in hydrostatic equilibrium, with the
thermal gas `mechanically' coupled to other components.  We define the
fraction of the pressure due to thermal gas as $f_g=f_g(r)$. Then the
pressure derived from X-ray data will be $nkT=f_gP$, where $P$ is the
true total pressure, $n$ and $T$ are the density and temperature
measured from X-ray data. If we further assume that only thermal gas
contributes to the mass density, then $\rho=\mu m_p n$ will be
correctly determined from the X-ray analysis. The temperature is also
assumed to be determined correctly.  Substituting $f_gP$ instead of
$P$ into eq.\ (\ref{eq:pot0}), we can easily express $\varphi_X(r)$
through the true potential $\varphi(r)$ and other gas parameters: 
\begin{eqnarray}
\varphi_X=\int{f_g d\varphi} -\frac{k}{\mu
m_p} \int{ T\frac{d\ln f_g}{dr} dr}.
\label{eq:phiu}
\end{eqnarray}
If $f_g$ does not depend on radius then obviously $\varphi_X=f_g
\varphi$. 

\begin{figure*}
\plottwo{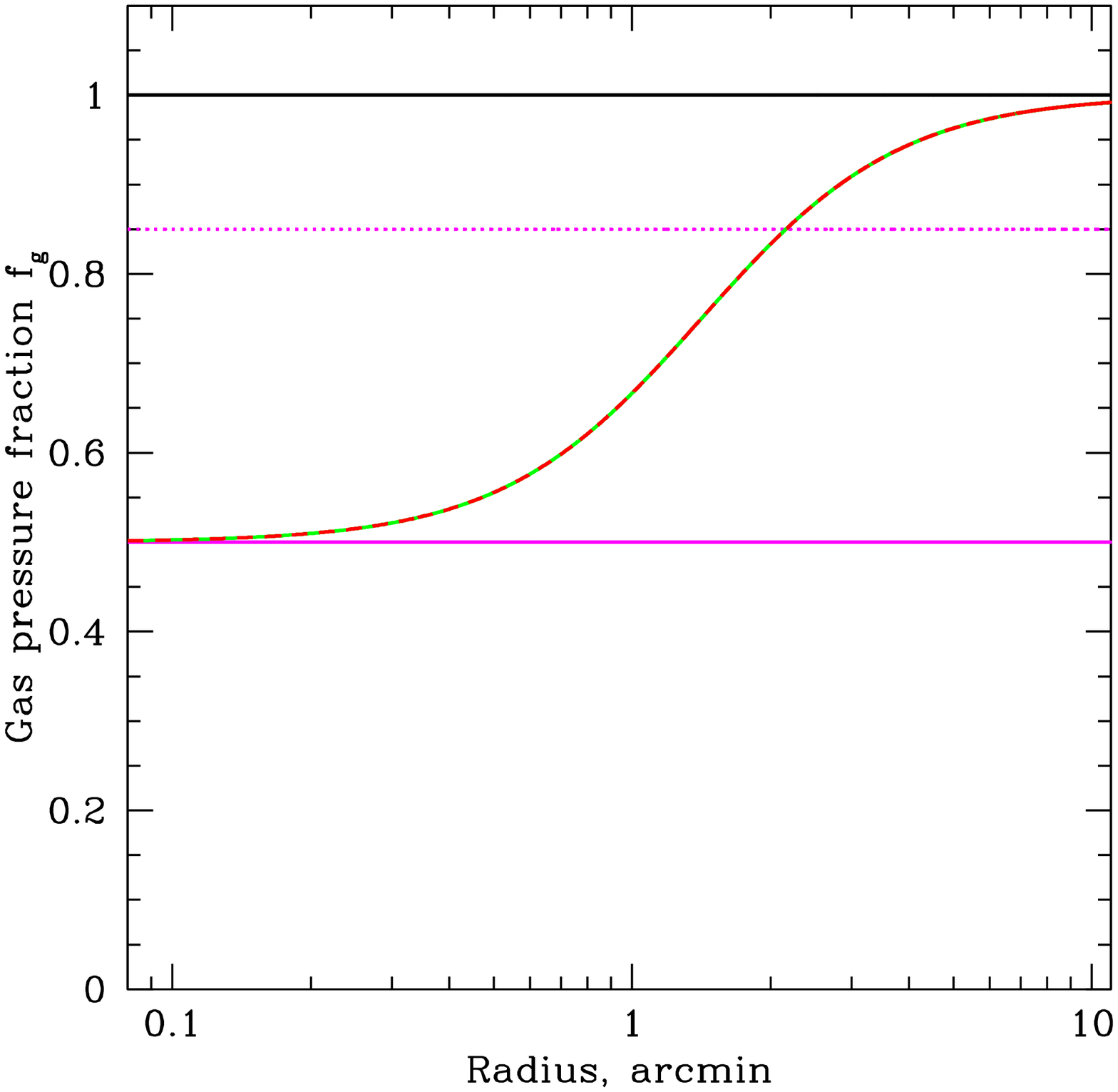}{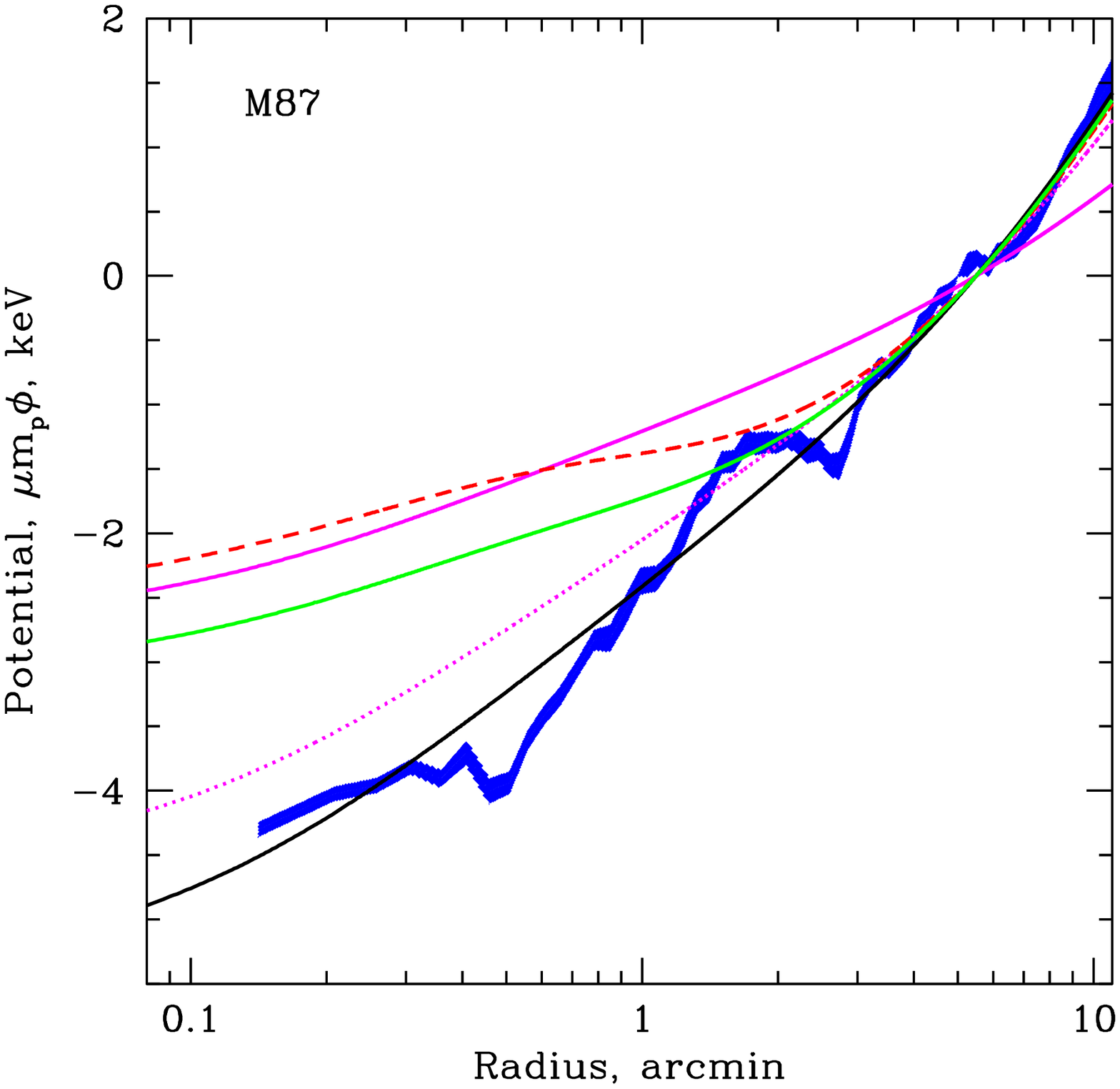}
\caption{{\bf Left:} Various model dependences of the thermal gas
fraction $f_g$ on radius discussed in Section \ref{sec:eff}. {\bf Right:} Expected potential $\varphi_X$
that would be derived from X-ray observations for each of the $f_g$
models shown in the left plot. The color coding is the same in both
plots.  We assumed that the true potential is the NFW2 model of
RK01. Such a potential would be derived from perfect X-ray data if
$f_g={\rm const}=1$ (thick black solid line).  Other $f_g$,
$\varphi_X$ pairs shown on both plots are: solid magenta line
represents $f_g={\rm const}=0.5$; dotted magenta line represents
$f_g={\rm const}=0.85$; red dashed line represents $f_g$ given by eq.\
(\ref{eq:fg}) and $\varphi_X$ from eq. (\ref{eq:phiu}); green solid
line represents $f_g$ given by eq.\ (\ref{eq:fg}) and $\varphi_X$ from
eq. (\ref{eq:phip}). Note, that the gas pressure fraction curves
for the two potentials in green and red (dashed) are identical
and they overlap in the left panel.
For comparison, the blue curve shows $\varphi_X$
derived from the Chandra observations of M87 (same curve as shown in
Fig. \ref{fig:pot1}). The comparison of black, magenta, red and green
lines in the right figure demonstrates the impact of the non-thermal
pressure on $\varphi_X$.  The measured $\varphi_X$ (blue line) agrees with
the optical profile (black line) better than any of the model curves,
suggesting that the fraction of nonthermal pressure is on average less
than is assumed in any of these models.
\label{fig:ngp}
}
\end{figure*}

Shown in Fig.\ \ref{fig:ngp} is $\varphi_X$ from eq. (\ref{eq:phiu}) for
$f_g={\rm const}=0.5$ and 0.85 (magenta solid and dotted lines respectively)
and for 
\begin{equation}
f_g=[1+(r/1')^2]/[2+(r/1')^2]
\label{eq:fg}
\end{equation} 
(red dashed line). The latter case illustrates the situation when gas
pressure dominates at large radii ($f_g=1$), but declines to 50\% of
the total pressure ($f_g=0.5$) in the core. In these calculations we
used $\varphi$ based on the NFW2 model of RK01 and the analytic
approximation of temperature given by eq. (\ref{eq:te}).  The curves
for $f_g=1$ and $f_g=0.85$ fit the X-ray data reasonably well, while
the other models are excluded.  For NGC 1399, the $\varphi_X$ and
$\varphi_{opt}$ closely follow each other. Values of $f_g$ smaller
than 0.9 (dotted magenta line in Fig.\ \ref{fig:ngc1399}) would lead
to a substantial disagreement between $\varphi_X$ and $\varphi_{opt}$
over the range of radii from few arcseconds to $\sim 2'$.

\subsection{Micro-turbulence}
\label{sec:mt}
If small scale and isotropic turbulent motions are present in the
medium, then the impact on $\varphi_X$ can be evaluated similar to the
case considered in the previous section. The quantity $1-f_g$
characterizes the contribution of turbulent motions to the pressure
(or energy density) of the medium.

\subsection{Bubbles of relativistic plasma}
\label{sec:patchy}
We now assume that the medium is not uniform. Let $f_g$ be the
volume fraction filled with thermal gas. The remaining
volume fraction is completely devoid of thermal gas and is occupied by
bubbles of cosmic rays and magnetic fields.  Both phases are assumed
to be in pressure equilibrium and coupled together. If, in performing
the X-ray
deprojection, one still assumes that the gas is uniform (as we did
above), the derived gas density will be lower than the true gas
density in the gas patches and higher than the mean volume density:
$\displaystyle
\rho_{obs}=\rho_{gas}\sqrt{f_g}=\rho_{mean}/\sqrt{f_g}$. The pressure derived
from the X-ray analysis will be modified accordingly. For this case, the
potential is:
\begin{eqnarray}
\varphi_X=\int{f_g d\varphi} -\frac{1}{2}\frac{k}{\mu
m_p} \int{ T\frac{d\ln f_g}{dr} dr}.
\label{eq:phip}
\end{eqnarray}
The X-ray potential 
$\varphi_X$ derived from eq. (\ref{eq:phip}) for
$f_g$ given by eq.\ (\ref{eq:fg}) is shown in Fig.\ \ref{fig:ngp} by
the green line.

We note here that a similar problem (the impact of bubbles on the
observed gas distribution) has been recently considered in Nusser \&
Silk (2007).
 
\subsection{Summary of the effects of non-thermal pressure components
  on $\varphi_X(r)$}

As discussed above we consider the following forms of non-thermal pressure
support as the most relevant: (i) cosmic rays and magnetic fields uniformly
mixed with the thermal gas (Section \ref{sec:uniform}), (ii) cosmic rays and
magnetic fields forming bubbles that are free of thermal gas (Section
\ref{sec:patchy}) and (iii) micro-turbulence in the thermal gas (Section
\ref{sec:mt}). For all these cases non-thermal pressure manifests itself as a
coefficient in the relation $d\varphi_X\approx f_g d\varphi$ if $f_g$ is
independent of radius. For $f_g$ varying with radius the dependence on
$f_g(r)$ is slightly more complicated (see eqs.\ \ref{eq:phiu} and
\ref{eq:phip}), with an additional term in $\phi_X(r_2)-\phi_X(r_1)$ roughly
equal to $-\displaystyle \frac{kT}{\mu m_p}\ln \frac{f_g(r_2)}{f_g(r_1)}$,
where $f_g(r_2)$ and $f_g(r_2)$ are the thermal gas pressure fractions at two
radii.  Therefore, the comparison of the change of $\varphi_X$ to the true
potential (if it is known) over the broad range of radii from $r_1$ to $r_2$
provides a simple and convenient way of estimating $f_g$---the contribution of
the non-thermal components to the total pressure---independent of the ``form''
of the non-thermal pressure. This applies also to combinations of
micro-turbulence, cosmic rays uniformly mixed with the thermal gas, and cosmic
rays contained in bubbles.

\subsection{Sound waves and shocks}
\label{sec:shock}
\begin{figure}
\plotone{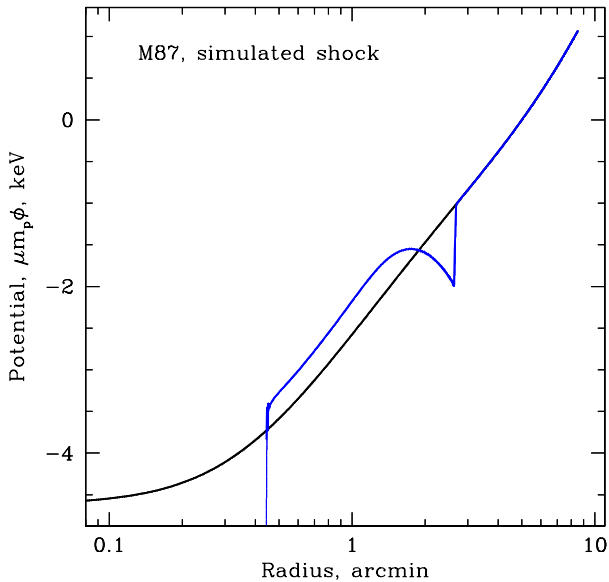}
\caption{Potential ($\varphi_X$ - blue line) derived using
eq. (\ref{eq:pot}) from simulated density and temperature profiles. In
these simulations a powerful explosion (with total energy a few
$10^{57}$ erg) at the center of the galaxy drove a shock through the
ICM. The data shown in the figure correspond to the  time
when the shock of Mach number $\sim 1.2$ is located $\sim 2'.7$ from
the nucleus, the observed radius of the strongest shock in M87. The true
potential is shown with the black line.  The deviations of
$\varphi_X$ predicted by the shock
model from the true underlying potential resemble the deviations between the observed $\varphi_X$ and
$\varphi_{opt}$ (e.g. blue curve in Fig.\ref{fig:ngp}), suggesting that
most of the wiggles seen in the observed $\varphi_X$ are due to the shock.
\label{fig:shock}
}
\end{figure}

The presence of sound and shock waves formally invalidates our assumption of
hydrostatic equilibrium. At any given location the potential
derived from X-ray data can be either underestimated or overestimated,
depending on the sign of the pressure gradient in the wave. Forman et
al. (2005, 2007) identified in the X-ray data a number of
quasi-spherical features around M87 which are plausibly caused by
shock waves produced by an unsteady outflow of relativistic plasma
from the black hole at the center of M87. These shock
candidates are located exactly where we see wiggles in the potential
(Fig.\ \ref{fig:ngp}). This is not surprising since, in a spherical
shock propagating from the center of the galaxy, the pressure first
increases sharply inwards at the front and then decreases in the
rarefaction region behind the shock. The potential derived from the
X-ray analysis (see eq. \ref{eq:pot}) will have a `dip' at the position
of the shock front and it will then recover from this dip over the
rarefaction region. To illustrate such behavior we simulated a
spherical shock in the atmosphere of M87 and calculated the potential
$\varphi_X$ which would be derived from X-ray data corresponding to
these simulations. Fig.\ \ref{fig:shock} shows the true potential,
assumed in the simulations, with the black line, while the blue line
shows $\varphi_X$ calculated for a Mach 1.2 shock located $2.7'$ from
the nucleus (see Forman et al. 2007,2008 for details of the shock
parameters and the simulations).

For weak shocks/sound waves which are well localized in space (e.g., weak
compression waves) $\varphi_X$ will be affected only in the region where the
wave is present, while $\Delta\varphi_X=\varphi_X(r_2)-\varphi_X(r_1)$ over a
range of radii from $r_1$ to $r_2$ which covers the entire perturbation will
not be affected. Therefore, weak and localized perturbations do not affect the
global change of $\varphi_X$. In the discussions below that
compare $\varphi_X$ and $\varphi_{opt}$ (Section \ref{sec:eval}) we focus on
radii, $r_1$ to $r_2$, away from the radii most strongly affected by the
shock.
 
\subsection{Non-luminous gas} 
In sections \ref{sec:uniform}, \ref{sec:mt} \ref{sec:patchy},
 we have assumed
that there is an `invisible' (in X-rays) component contributing to the
pressure. As a result, the derived potential $\varphi_X$ was flatter
than the true potential. We can also assume that there is an X-ray
`invisible' gas component which contributes to the mass density and
has a small filling factor. A cool gas phase with temperature below
`observable' X-ray values but coupled to the X-ray emitting phase
(e.g., by magnetic fields) would be the simplest example. The
$\varphi_X$ derived from equation (\ref{eq:pot0}) will be simply:
\begin{eqnarray}
\varphi_X=\int{f_m d\varphi},
\label{eq:phicg}
\end{eqnarray}
where $f_m=\rho/\rho_X$ is the ratio of total mass density to the mass density
derived from the X-ray analysis. In this case $f_m>1$ and the X-ray derived
potential will be steeper than the true potential. Although we note this
possibility here, for the remainder of the discussion below, we assume that
this ``non-luminous gas'' component does not affect M87 and NGC1399. Otherwise
it will be difficult to disentangle the contribution of such a component from
that of non-thermal components which affect $\varphi_X$ in the opposite
direction. In the absence of any ``non-luminous gas'' component, we can place
constraints on the non-thermal pressure as we discuss below.

\section{Discussion}
\subsection{Evaluation of the non-thermal pressure contribution
$f_g$}
\label{sec:eval}
\begin{figure*}
\plottwo{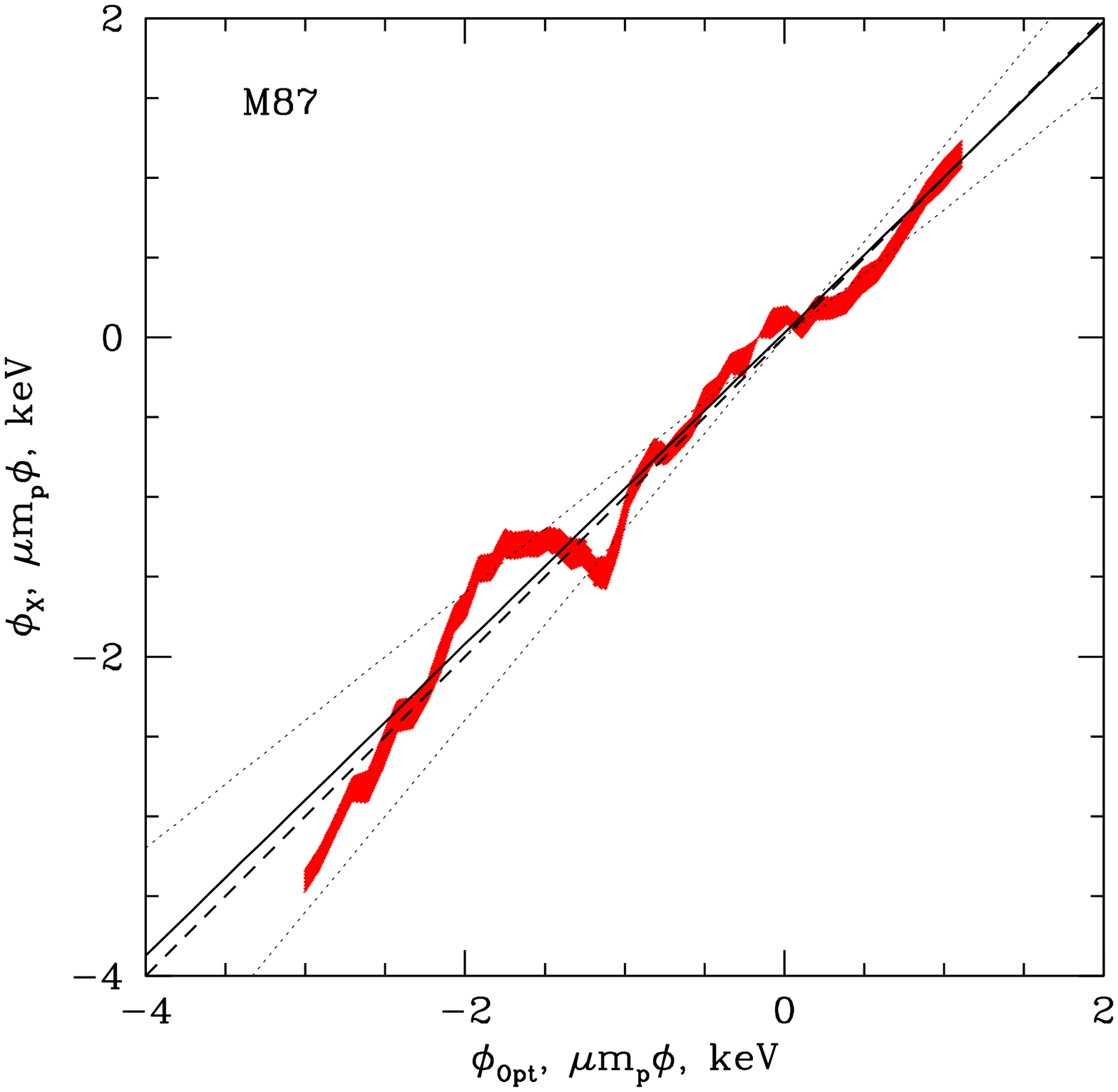}{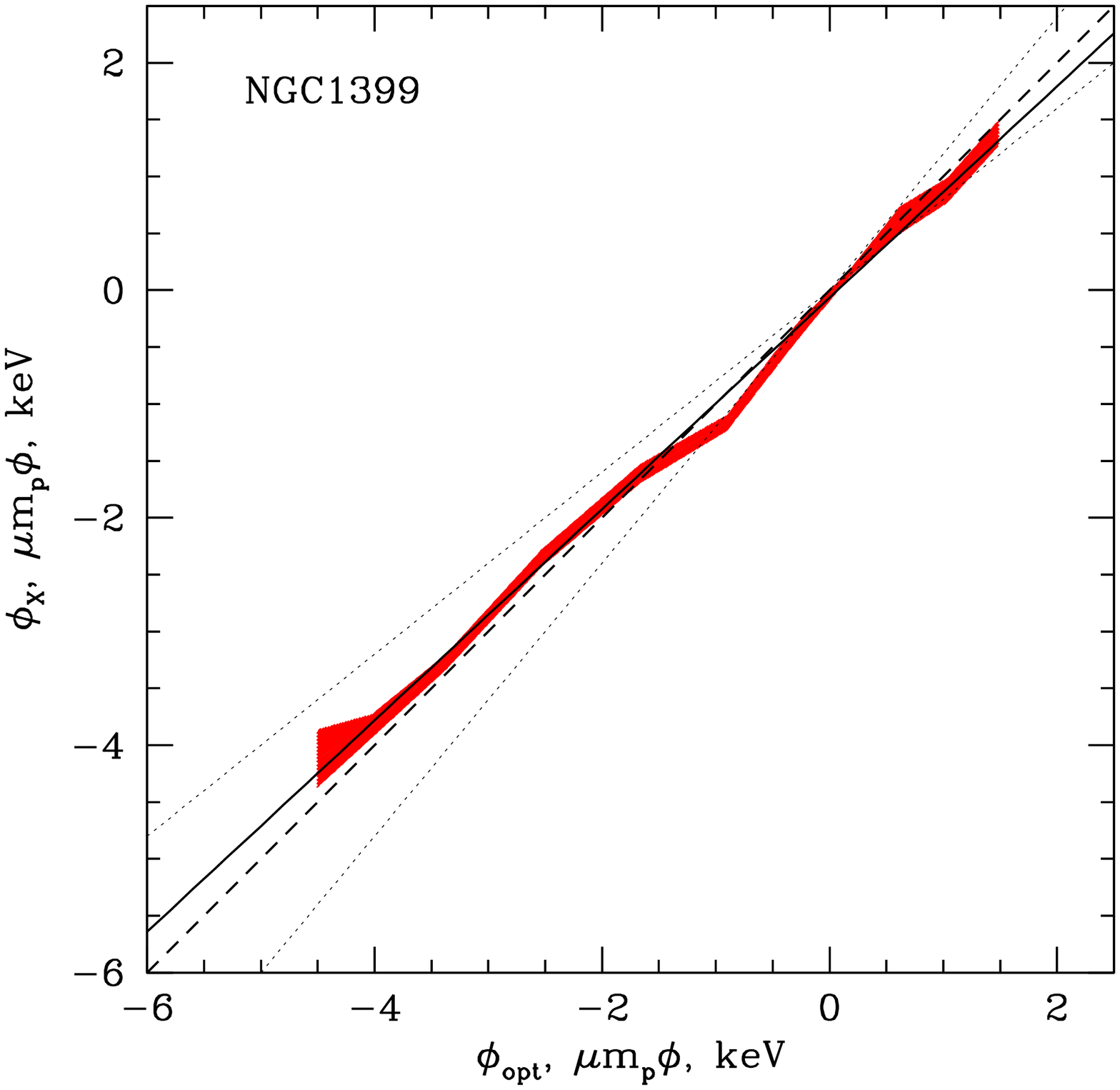}
\caption{ Potential derived from X-ray data 
plotted versus the
potential derived from optical data for M87 and NGC 1399 (red shaded
areas). The thick solid line is a formal linear fit
$\varphi_X=a \varphi_{opt}+b$. In M87 the value of $a$ is 0.975
for NWF2 model of RK01 and 0.89 for WT06 model. In NGC 1399  $a=0.93$
for Kronawitter et al. 2000 model. The thick dashed line shows the dependence 
$\varphi_X=
\varphi_{opt}$, while thin dotted lines correspond to $a$=0.8 and
1.2. For both objects,  values of $a$ close to unity imply that
the fraction of non-thermal pressure is $\lesssim$0.1.
\label{fig:xo}
}
\end{figure*}

 In Fig.\ \ref{fig:xo} the
potential derived from X-ray data is plotted against the potential
derived from optical data for M87 and NGC 1399 (red shaded areas). If
the assumptions used when deriving the potentials are correct, then
one expects to find $\varphi_X=\varphi_{opt}$ (shown with the dashed
lines). If cosmic rays, magnetic fields or micro-turbulence contribute
to the pressure in the X-ray emitting gas, then the change of
$\varphi_X$ between two radii will be smaller than the change in
$\varphi_{opt}$, i.e. $\Delta\varphi_X=a\Delta\varphi_{opt}$, where
$a<1$. In the simplest case of a constant $f_g$, there
is a linear relation between the two potentials $\varphi_X=a
\varphi_{opt}+b$, where $a=f_g$. As shown in sections \ref{sec:uniform} and
\ref{sec:patchy}, if $f_g$ depends on radius then the relation between
the potentials can be more complicated. If $f_g$ is not constant, but
decreases towards small radii, i.e. the role of non-thermal pressure
increases towards the center, then  $\varphi_X$ will be even 
shallower than is prescribed by the relation
$\varphi_X=f_g\varphi_{opt}+b$ (see eq.  \ref{eq:phiu} and
\ref{eq:phip}). Therefore, we can use the relation $\varphi_X=a
\varphi_{opt}+b$, corresponding to the assumption of constant $f_g=a$,
to provide a conservative upper limit on the
contribution of non-thermal pressure $1-f_g$ over the observed range of
radii. This upper limit corresponds to an averaged non-thermal pressure
contribution over the observed range of radii. The averaging is
 done with effective ``weights'' corresponding to the values of
the optical potential at each radius. For example, for an isothermal
(logarithmic) potential the weight of the radial range from $r$ up to
$r+\Delta r$ is proportional to $\displaystyle \ln \frac{r+\Delta r}{r}$.  

A linear fit
to the observations of the form $\varphi_X=a \varphi_{opt}+b$ is shown
in Fig.\ \ref{fig:xo} with the thick solid line. For NGC1399, $a=
0.93$. For M87 $a= 0.89$ if the WT06 model is used and $a= 0.975$ for
the NFW2 model of RK01. The very closeness of $a$ to unity in the
latter case (NWF2 model) is surely a coincidence given the obvious
wiggles in $\varphi_X(r)$ (Fig.\ \ref{fig:xo}).  For the two other
(less probable) models of RK01 - NFW1 and NFW2 the values of $a$ are
0.96 and 1.21 respectively.  By repeating the fit four times using
independent quadrants of the X-ray data (see Section 7.2) we conclude
that the statistical uncertainty in these slopes is $<0.04$.  Taken at
face value, these results mean that the fractional non-thermal
contribution to the pressure amounts to 7\% for NGC1399 and 2.5-11\%
(depending on the $\varphi_{opt}$ model) for M87. These values
characterize the non-thermal contribution averaged over the range of
radii from $0'.7$ to $12'$ in M87 and from $3''$ to $5'$ in NGC 1399.
Given that systematic uncertainties are present in the data (e.g., due
to the shock in M87) it is difficult to provide a precise estimate of
the uncertainty associated with this value. If one adopts a more
conservative approach of upper limits (rather than measurements) on
the non-thermal pressure components, then these limits are $\sim10$\%
and $\sim20$\% for NGC 1399 and M87 respectively (see Fig.\
\ref{fig:xo}).

Fig.~\ref{fig:bias} (top panel) shows that, for temperatures characteristic of
M87 and NGC1399, fitting multi-phase plasma spectra with a single-temperature
model will result in $\varphi_X$ biased towards shallower values (smaller
$a$). Therefore this ``multi-phase bias'' has the same sign as the non-thermal
pressure and can only lead to an overestimation of the non-thermal pressure
fraction; thus the upper limits given in the preceding paragraph remain
valid. In addition, based on the comparison of spectral fitting in two energy
bands (Fig. \ref{fig:dspec}), we do not expect this bias to be strong,
at least outside the innermost $0'.4$ in M87.

Errors in the optical potential profile could also contribute to the
observed difference in potentials.  In both galaxies, uncertainties
arise from the assumption of spherical symmetry, the restriction to a
particular family of parametric halo potentials in the dynamical
modeling, and possible systematic errors in deriving the line-of-sight
velocity distribution from the observed spectra. These uncertainties
are difficult to estimate  precisely.  Confidence limits
  within the parametric potentials used for NGC 1399 are given in
  Fig.~\ref{fig:ngc1399} and Section~\ref{sec:ngc1399}.  Thomas et
al. (2007) have investigated the uncertainties in the inferred
circular velocity curves for a sample of Coma cluster ellipticals
which they studied with axisymmetric dynamical models 
  and in a wider range of potentials.  In these galaxies, the central
mass density is dominated by the luminous matter, independent of the
dark matter halo profile used; thus the circular velocities inside
$\simeq 0.5R_e$ are accurately determined, within $\simeq 5\%$. At two
effective radii, the uncertainties in the inferred circular velocities
for their galaxies are $\simeq 15\%$.  The measurement errors in the
observed kinematics for NGC 1399 (Saglia et al. 2000) are smaller than
in the Coma ellipticals. However, NGC1399 observations are along only one long
slit. Hence, we would expect somewhat smaller uncertainties in NGC
1399.   Again the error in the potential is smaller,
  since the potential is obtained by the integral of the circular
  velocity curve. However, note that differences $\phi_{opt}-\phi_{true}$,
in the inferred optical potential profiles are likely to be more
slowly varying functions of radius than the variations in
$\phi_X-\phi_{opt}$ seen in Fig.~\ref{fig:ngc1399}.

The
$\varphi_X-\varphi_{opt}$ difference characterizes the combined
contribution of all  non-thermal pressure components and the
modelling uncertainties of the respective datasets. Under the
assumption that modelling uncertainties do not dominate and that the
contributions of individual components of the non-thermal pressure all
have the same sign, the measured values can be converted to upper
limits. For example, fractions of non-thermal pressure support of 
  10\% in M87 and NGC 1399 translate into upper limits on the
magnetic field of $\sim$16 and $\sim$10 $\mu$G respectively (evaluated
at a distance from the galaxy center of $2'$ in M87 and $1'$ in NGC
1399).

\subsection{Non-spherical models}
\label{sec:nonsph}
We have assumed that the gravitational potential in both M87 and NGC 1399 is
spherical. The ellipticity of the optical isophotes of M87 is near zero at the
center, rising to 0.4 at $10'$, while the ellipticity of the X-ray isophotes
outside $5'$ is 0.1--0.15, with the same position angle (B\"ohringer et
al. 1997) (see Fig.\ref{fig:image_m87}). The ellipticity of the optical
isophotes of NGC 1399 is about 0.1 out to $10'$ (Dirsch et al. 2003). The X-ray
image of NGC1399 looks more disturbed (see Fig. \ref{fig:image_ngc1399})
than the optical image and has an ellipticity of 0.34 at the effective radius
of the galaxy (Diehl \& Statler 2007). The shapes of the X-ray emitting gas
and the stellar distribution need not be the same  in the
common potential, since the
velocity-dispersion tensor of the stars is not necessarily isotropic.

\begin{figure*}
\plottwo{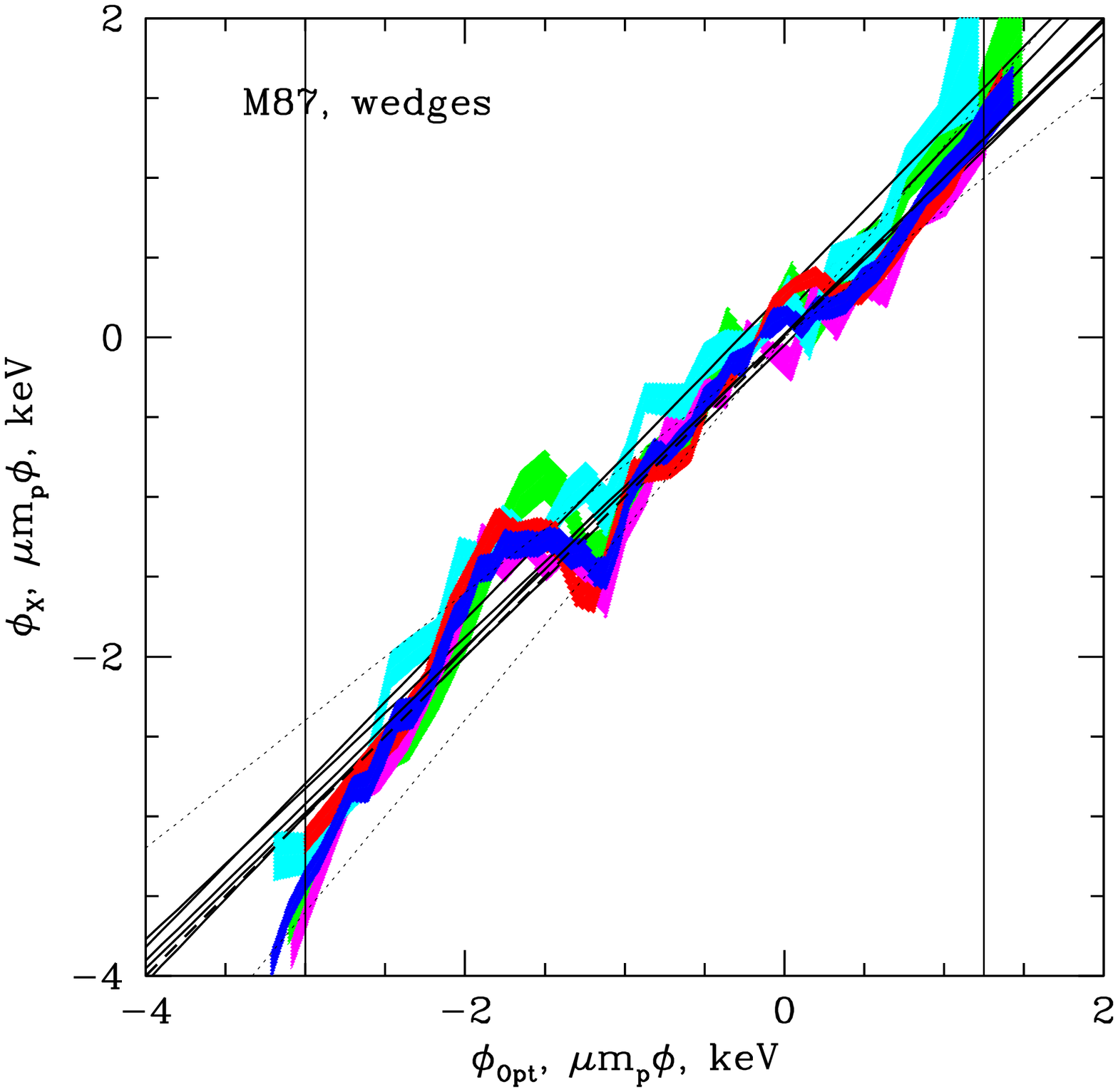}{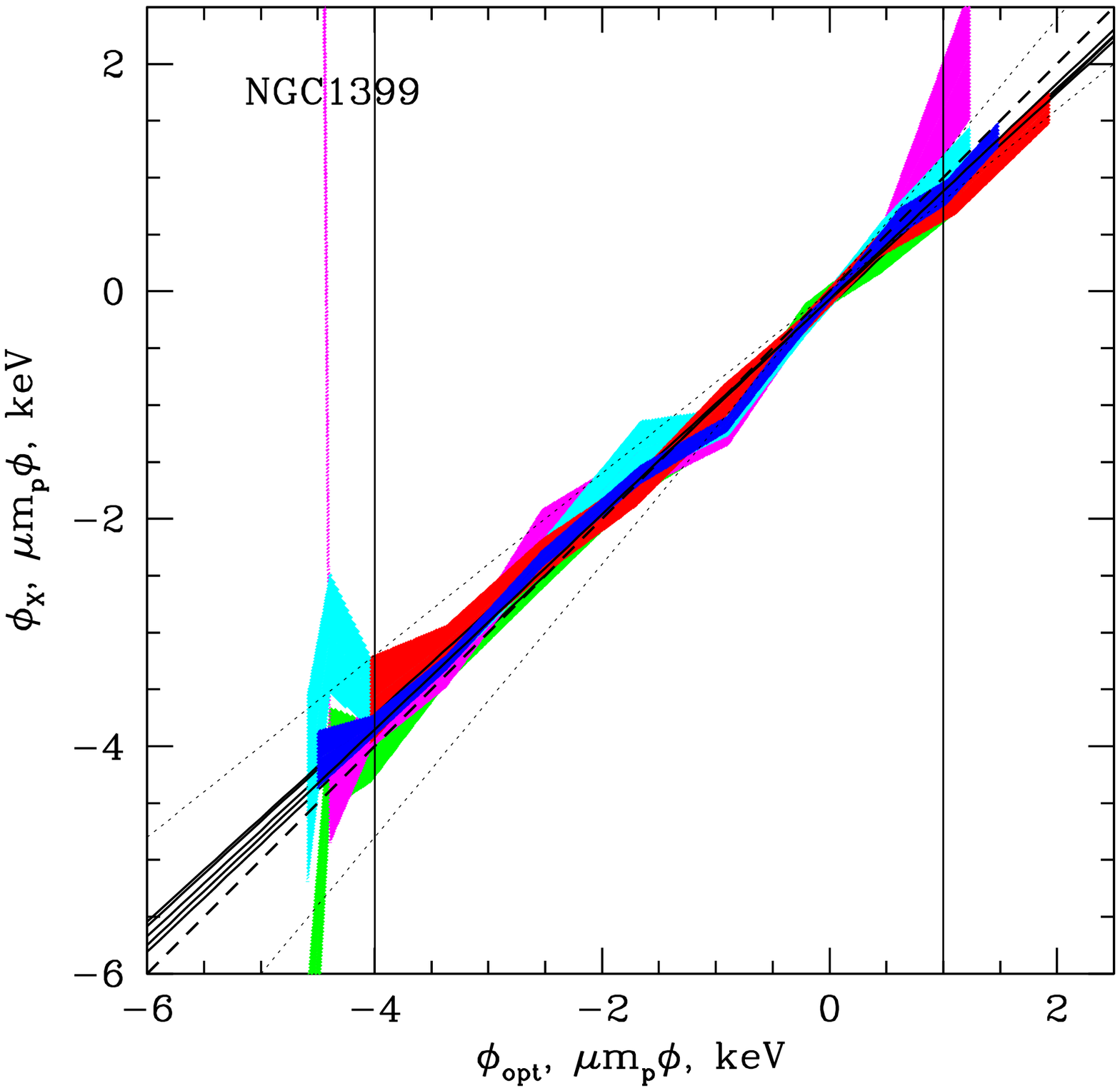}
\caption{Potentials derived from X-ray data for M87 (left) and NGC1399 (right)
in four 90\deg ~wedges (green, magenta, cyan and red for NW, NE, SE and SW
wedges respectively), plotted against the optically derived potentials (RK01
for M87 and Kronawitter et al. 2000 for NGC1399). The centers of the wedges
coincide with the optical centers of the galaxies. The blue curves in each
plot show the X-ray potentials derived from the full 0-360\deg data. Black
solid lines show the linear least-square fits $\varphi_X=a \varphi_{opt}+b$
for each wedge and for the complete dataset. The innermost and outermost bins,
which have large statistical uncertainties, were excluded from the fit (thin
vertical lines). As in Fig.\ \ref{fig:xo}, the thick dashed line shows the
dependence $\varphi_X= \varphi_{opt}$, i.e. $a=1$, while the thin dotted lines
correspond to $a$=0.8 and 1.2. Clearly the slopes $a$ derived for
individual independent wedges agree well with each other, with
root-mean-square deviations of 4.5\% for M87 and 1.5\% for NGC1399. The
deviations characterize the overall uncertainties in $a$ that are caused by
the apparent asymmetries in the X-ray images and by statistical errors.
\label{fig:wedges}
}
\end{figure*}

Thus both galaxies exhibit a modest level of non-axisymmetry in their X-ray
and optical isophotes.  We can assess the
impact of these asymmetries on the X-ray derived potential by using the data
in individual (independent) wedges and deriving the corresponding X-ray
potentials as shown in Fig. \ref{fig:wedges}. Within each 90\deg ~wedge,
centered at the optical center of the galaxy, we assume that the gas
properties depend only on the distance from the center and we repeat our
previous analysis up to fitting the $\varphi_X=a \varphi_{opt}+b$ relation.
The innermost and outermost bins, which have large statistical uncertainties,
were excluded from the fit (as shown by thin vertical lines in
Fig.~\ref{fig:wedges}). The values $a_{\rm M87, NWF2}=\{1.04,0.98,1.03,0.95\}$ and
$a_{\rm NGC1399}=\{0.95, 0.93, 0.92, 0.95\}$ were found for the NW, NE, SE and
SW wedges in the two galaxies. If WT06 model is used for M87 then
$a_{\rm M87, WT06}=\{0.93,0.86,0.89,0.87\}$. 
The root-mean-square deviations of $a$ from the
mean values are 3-4\% for M87 (depending on the optical profile used) and 1.5\% for NGC1399. These modest deviations
characterize the total uncertainty associated with the apparent
non-axisymmetry in the X-ray isophotes of these two galaxies.  Note that the observed scatter in slope between
the four quadrants also represents an upper limit to the statistical
uncertainties in the slope, although for these datasets we believe that the
scatter is mostly due to non-axisymmetry rather than statistics.

An additional uncertainty, probably of a few per cent, is related to the
freedom of choosing the range of radii used for fitting the $\varphi_X=a
\varphi_{opt}+b$ relation. This is more serious for M87 where the presence of
a shock clearly makes the results more sensitive to the choice of the radial
bins.

Another concern is that the actual deviations from spherical symmetry
may be larger than the observations indicate, if the galaxy is an
approximately axisymmetric system that is viewed pole-on.  The effects
of a non-spherical potential on X-ray estimates of mass or potential
have been modelled by Piffaretti, Jetzer \& Schindler (2003) for
an ellipsoidal $\beta$-model (see also Gavazzi 2005 for the analysis of
axisymmetric NFW profiles). Although Piffaretti et al. were concerned
with clusters of galaxies, their results apply to galaxies as
well. They found that if (i) the galaxy potential is axisymmetric and
viewed pole-on and (ii) its mass is estimated from the temperature and
distribution of hot gas with the assumption that the potential is
spherical; then, the derived mass within a spherical radius is
typically in error by only a few per cent, if the axis ratio of the
potential is varied within plausible limits (typically $\pm 40\%$).
  We note that for the model considered by Piffaretti et
al. (2003) (ellispoidal $\beta$-model distribution of
isothermal gas) in the limiting case of the radius much larger than
the $\beta$-model core radii the deprojected potential/mass profiles are
exact for any orientation of the galaxy/cluster with respect to the
line of sight (see Appendix \ref{sec:anonsph}). While the above
assumptions (e.g. isothermality) are not in general met, this result
suggests that the effect of triaxiality can be small. We provide a more
quantitative analysis of the triaxiality on real systems in future
work.

The effects of a non-spherical potential on optical estimates of mass are more
difficult to model. The simplest approach is based on the tensor virial
theorem (e.g.,  Binney \& Tremaine 2008). If (i) an axisymmetric one-component
self-gravitating stellar system is viewed pole-on; (ii) the
density of stars in the system is constant on spheroids of
eccentricity $e$; and
(iii) the mass of the system is estimated using the virial theorem for
spherical systems; then the derived mass will exceed the actual mass by a
factor $f$, where
\begin{equation}
f(e)=\left\{\begin{array}{ll}
     3{\displaystyle 1-e^2\over \displaystyle e^2}\left({\displaystyle
     1\over\displaystyle \sqrt{1-e^2}}-{\displaystyle \sin^{-1}e\over
     \displaystyle e}\right) &
     \mbox{oblate systems} \\
     3{\displaystyle \sqrt{1-e^2}\over \displaystyle e^2}\left({\displaystyle
     1\over\displaystyle 2e}\log{\displaystyle 1+e\over \displaystyle 1-e}-1\right) &
     \mbox{prolate systems},
         \end{array}\right.
\end{equation}
where $e=\sqrt{1-b^2/a^2}$ and $b/a$ is the axis ratio. For a plausible range
of axis ratio, say $0.5<b/a<1$, $f$ varies only between 0.79 (most oblate) and
1.04 (most prolate), suggesting that the error in the optical potential due to
asphericity is likely to be small. 

 It
should also be stressed that the probability that either of the studied
galaxies is strongly oblate or prolate and viewed pole-on is rather low. The
distribution of ellipticities in a sample of brightest cluster ellipticals
similar to M87 and NGC1399 was studied by Porter et al. (1991). They modelled
the distribution of axis ratios $q$ as a Gaussian with mean $\mu$ and
dispersion $\sigma_e$, and found $\mu=0.66$, $\sigma_e=0.09$ if the galaxies
were assumed to be oblate, and $\mu=0.69$, $\sigma_e=0.09$ if the galaxies
were prolate (at a radius of 15 kpc, corresponding to $3'.2$ in M87 and $2'.6$
in NGC 1399). The apparent axis ratios in M87 and NGC1399 at this radius are
0.2 and 0.1 respectively. The conditional probability distribution of the
intrinsic axis ratio, given the apparent axis ratio, can be computed from
these data. The probability that the intrinsic axis ratio in the stellar light
distribution is as small as 0.5 is only 2\% if the galaxies are oblate and 6\%
if they are prolate. The potential due to the stars is, of course, more nearly
spherical than the distribution of the stars themselves. 

Finally, we point out that substantial errors due to asphericity are
likely to affect the optical and X-ray estimates of the potential in
different ways. The close agreement between the X-ray and optical
potentials, seen in Figs.~\ref{fig:pot1} and \ref{fig:ngc1399}, would
be an unlikely coincidence if any asphericity were significantly
affecting the derived potentials.

\subsection{Is the small non-thermal component consistent with AGN-heated cluster cool cores?}
 
The central region of M87 is a classic example of a cooling flow -- a
region where gas radiative losses are high and an external source of
energy is required if one wants to have a quasi-steady state without
large net cooling of the gas, as indicated by the recent XMM-Newton
RGS data (e.g., Peterson et al. 2003). A plausible source of energy
for gas re-heating is the activity of the central black hole, mediated
by the outflow of relativistic plasma. Actual gas heating might be due
to dissipation at shock fronts (e.g., David et al. 2001) or in sound
waves (if the medium is viscous, has appreciable thermal conductivity
or is strongly non-uniform, e.g., Fabian 2003, 2005;  Ruszkowski et
al. 2004a, 2004b; Heinz \& Churazov 2005) or  through an intermediate
step of generating gas turbulence, which eventually dissipates its
energy (e.g., Churazov et al. 2001, 2002). We now test whether our
results on the non-thermal forms of pressure are consistent with this
picture.

First, we directly see bubbles of relativistic plasma (as radio
bright regions and as `cavities' in X-rays). If the M87 gas is in a
quasi-equilibrium state with cooling losses, on average compensated by
heating, then the mechanical power of the outflow should be related to
the gas cooling losses. As argued in Churazov et al. (2001), Begelman
(2001), and Nulsen et al. (2006), the efficiency of energy transfer from
a relativistic outflow to the gas is high, when the outflow subsonically
crosses regions with large pressure gradients.  In this case, one
can simply assume that the power of the outflow is equal to the
cooling losses. The total energy of the relativistic plasma bubbles
in the region can then be estimated from the balance of powers as
$\displaystyle \mathcal{E}/t_{cross}\sim
\mathcal{E}_{thermal}/t_{cool}$, where $\mathcal{E}_{thermal}$ is the
gas thermal energy, $t_{cool}$ is the gas cooling time and $t_{cross}$
is the time each bubble spends inside the region where it deposits
a significant fraction of its enthalpy.  For example, consider a 10 kpc
region around M87. For an electron density $n_e=2 \times 10^{-2}~{\rm
cm^{-3}}$ and a temperature $T_e=1.7~{\rm keV}$, the cooling time
$t_{cool}\sim 7\times 10^8~{\rm yr}$.  On the other hand, the rise velocity
of large buoyantly driven bubbles is $\varsigma c_s$ where $c_s$ is
the sound speed and $\varsigma$ is a factor of order 0.5
(e.g., Churazov et al. 2001). Thus, $t_{cross}\approx3\times 10^7~{\rm yr}$
for the 10 kpc region. Therefore, bubbles of relativistic plasma should, on
average, account for a fraction $t_{cross}/t_{cool}\sim 0.05$  of the total
energy (or volume). This is consistent within a factor of 2
with the results we obtained from Fig.\ \ref{fig:xo}. 
Note that since the electron density in a typical cooling flow
(e.g., in M87) scales approximately as
$n_e\propto r^{-1}$, then $t_{cool}\propto 1/n_e \propto r$ and
$t_{cross}\propto r$. Therefore the above estimate of
$t_{cross}/t_{cool}$ is only weakly sensitive to the radius used to evaluate
all quantities.
  
We can now check whether our limits exclude turbulence as an intermediate
reservoir of energy in the process: moving bubbles $\rightarrow$
turbulence $\rightarrow$ viscous dissipation and gas heating as
suggested by Chandran (2004) and Rebusco et al. (2005, 2006). 
In this case, the energy
dissipation rate of the turbulent motions should be approximately
equal to the gas cooling losses:
\begin{eqnarray}
C~\mu m_p n~\frac{v_{turb}^3}{l_{turb}} \approx \frac{\frac{3}{2}nkT}{t_{cool}},
\label{eq:turb1}
\end{eqnarray}
where $v_{turb}$ and $l_{turb}$ are the characteristic velocity and
spatial scale of energy containing eddies, $n$ is the gas particle density 
and $C$ is a dimensionless constant of the order of
unity. According to a recent compilation by Dennis \& Chandran (2005),
$C\simeq 0.42$. If the energy in
the micro-turbulence is constrained to be less than a fraction $\xi$ of
the thermal energy, then 
\begin{eqnarray}
\frac{1}{2}\mu m_p n~v_{turb}^2 \lesssim \xi \frac{3}{2}nkT,
\label{eq:turb2}
\end{eqnarray}
and then
\begin{eqnarray}
\frac{v_{turb}}{l_{turb}} \gtrsim \frac{1}{2C}\frac{1}{\xi} \frac{1}{t_{cool}}.
\label{eq:turb3}
\end{eqnarray}
Therefore $\displaystyle v_{turb}\gtrsim 20 \left( \frac{l_{turb}}{{\rm
1~kpc}}\right)\left( \frac{\xi}{0.15}\right)^{-1}~{\rm km~s^{-1}}$. This condition is likely satisfied in
the gas around M87 (if it is indeed turbulent), since the characteristic
size of individual bubbles is of order 1--3 kpc and their velocities are
expected to be significantly higher, a few $10^2~{\rm km~s^{-1}}$
(Churazov et al. 2001).

\subsection{The history of the ICM enrichment with cosmic rays}
 
\begin{figure}
\plotone{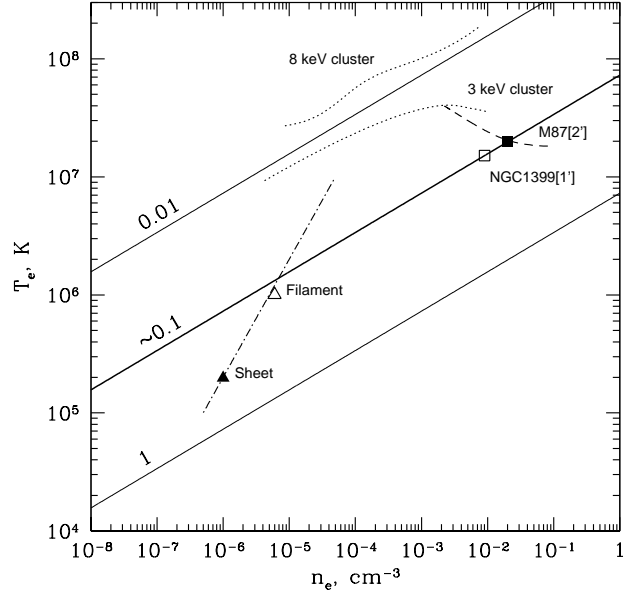}
\caption{Constraints on the energy density of cosmic rays during the
evolution of gas lumps presently seen in the cores of M87 and
NGC 1399. The ratio $\varepsilon'_{cr}/\varepsilon'_{gas}$ is
constant along the solid lines. If the gas had in the past a density
and temperature along the thick solid line, then at that time the
constraint on energy density of cosmic rays is the same as now:
$\varepsilon'_{cr}/\varepsilon'_{gas} \lesssim 0.1$. Thin
solid lines show the parameters for which present day constraints
transform into $\varepsilon'_{cr}/\varepsilon'_{gas}$ limits
of 0.01 and 1, respectively. For comparison we show a few typical
states of the IGM/ICM, ranging from warm gas in sheets and filaments
(triangle symbols and the dash-dotted line) in the large scale
structure to shock-heated gas (3 and 8 keV) in galaxy clusters (dotted curves).  The
dashed line shows the locus of $(n_e, T_e)$ points taken from radially
averaged profiles of the ICM in M87 at distances from $0'.5$ to $5'$
from the galaxy center.
\label{fig:crays}
}
\end{figure}

Our constraints on the non-thermal pressure are consistent with
the current paradigm of re-heating of gas in ``cooling flows'',
which assumes a balance between radiative cooling and mechanical
heating of the gas. On the other hand, our conclusion that the
fraction of the energy density in the form of cosmic rays and magnetic
fields is $\lesssim$10-20\% provides an important constraint on the
history of the ICM. In particular, these limits apply to the mixing of
thermal plasma with relativistic plasmas, as well as the generation
of cosmic rays by shocks in the bulk of the ICM. Generally, it is
believed that the energy density of cosmic rays is dominated by
protons rather than electrons and the particles with the lowest
$\gamma$ provide the most important contribution to the energy
density. The lifetime of a transrelativistic proton (Lorentz factor of
2) with
respect to Coulomb energy losses in a cluster plasma with electron
density of $10^{-2}~{\rm cm^{-3}}$ is roughly 5 Gyr. Therefore, the
energy density of relativistic protons is largely conserved, except
for adiabatic losses, during the lifetime of the cluster
(transrelativistic electrons have of course much shorter
lifetimes). This implies that our constraints on the energy density of
cosmic rays in the cores of M87 and NGC 1399 today actually limit the
presence of relativistic protons during the entire history of the same
gas lump, provided that relativistic particles are not able to diffuse
out of the lump. The constraints on cosmic ray energy density in the
past may be even tighter, if cosmic rays were originally deposited into
a gas lump which had much higher entropy than is observed today and
has lost the excess entropy by cooling.  Indeed, if relativistic
protons experience only adiabatic losses, then their energy density
changes with gas density as:
\begin{eqnarray}
     \varepsilon'_{cr}&=&\varepsilon_{cr} \left(
          \frac{\rho'_{gas}}{\rho_{gas}}\right)^{\gamma_{cr}},
\end{eqnarray}
where $\rho_{gas}$ and $\varepsilon_{cr}$ denote the thermal gas
density and the energy density of cosmic rays observed now;
$\gamma_{cr}=4/3$ is the adiabatic index of cosmic rays. All
quantities measured at a different time are denoted with a prime. The thermal energy is by definition
\begin{eqnarray}
\varepsilon'_{gas}&=&\varepsilon_{gas}\left(
\frac{\rho'_{gas}}{\rho_{gas}}\frac{T'_{gas}}{T_{gas}}\right).  
\end{eqnarray}
Thus
\begin{eqnarray}
\frac{\varepsilon'_{cr}}{\varepsilon'_{gas}}&=&\frac{\varepsilon_{cr}}{\varepsilon_{gas}}\times
\left(\frac{\rho'_{gas}}{\rho_{gas}}\right)^{\gamma_{cr}-1}
\frac{T_{gas}}{T'_{gas}}.
\end{eqnarray}
Therefore, a constraint on the energy density of cosmic rays now
$\varepsilon_{cr}/\varepsilon_{gas}\lesssim 0.1$,
for M87 and NGC 1399, can be transformed into constraints on
$\varepsilon'_{cr}/\varepsilon'_{gas}$ in the same gas
lump. This is illustrated in Fig.~\ref{fig:crays}. Solid and open
squares show the gas density and temperature observed in the cores of
M87 and NGC 1399 today. Along the solid lines ($T\propto
\rho^{\gamma_{cr}-1}$), the ratio
$\varepsilon'_{cr}/\varepsilon'_{gas}$ is
constant. Therefore, if, in the past, the gas lump we see now in the
core of M87 or NGC 1399
had a density and a temperature along the thick solid line, then
the constraint on the energy density of cosmic rays would be the same as
now: $\varepsilon'_{cr}/\varepsilon'_{gas} \lesssim
0.1$. The two thin solid lines show the parameters for which present day
constraints transform into 
$\varepsilon'_{cr}/\varepsilon'_{gas}$ limits of 0.01 and 1
respectively. For comparison, we show a few typical states of the
IGM/ICM, ranging from warm gas in sheets and filaments (triangle
symbols and the dash-dotted line) in the Large Scale Structure to
shock heated gas (3 and 8 keV) in galaxy clusters (dotted curves). The cluster data
are taken from simulations of Nagai, Vikhlinin \& Kravtsov (2007) for
$z=0$. The data for 3 keV and 8 keV clusters were used and the radial
profiles plotted in Fig.\ \ref{fig:crays} correspond to distances from
$\sim$70 kpc to $\sim$2 Mpc from the cluster center.  The main
conclusion of this simple exercise is that if the gas in the core of
M87 or NGC 1399 were in the past shock heated above the upper solid
line, the energy density of cosmic rays in the gas, at that time,
was less than 1-2\% of the thermal energy density. We emphasize again that
this conclusion is based on the assumption that cosmic rays dominating
the energy density (presumably low energy protons) are not able to
diffuse out of the gas lump and they suffer only  adiabatic
losses. Allowing diffusion would essentially imply that our upper
limits are applicable only to the moment of observation.

Gamma-ray observations of clusters (Reimer et al. 2003) do limit the
presence of cosmic rays, in particular protons, to less than
$\sim$20\% of the gas thermal energy (e.g., Pfrommer \& Ensslin 2004). 
It is interesting that Brunetti et al. (2007) have recently derived an
upper limit on the contribution of relativistic protons to the gas
pressure based on radio upper limits for clusters
without radio halos. While the actual limit on protons is sensitive to
the assumed strength of the magnetic field, for many clusters the
upper limit falls
below 1\% (for $>\mu G$ fields) of the thermal pressure measured from
the X-ray gas.

\subsection{Clusters of galaxies}

X-rays can be used to constrain mass/potential profiles of elliptical
galaxies as long as the galaxy is gas rich and the contribution of
individual stellar-mass systems to the X-ray emission (e.g.,
Revnivtsev et al. 2007) can be neglected. This implies that the X-ray
analysis described in this paper can be applied to massive ($\gtrsim
10^{11}~M_\odot$) systems. The most massive ellipticals often dwell at
the centers of groups/clusters (galaxies like NGC1399 and M87). For
these systems, X-ray observations can be used over a broad range of
radii, from a fraction of the central galaxy effective radius to
distances almost comparable to the cluster virial radius. The optical
data on stellar kinematics are mostly useful up to 1-2 $R_{e}$. This
range of radii can be extended to a few $R_{e}$ with tracers such as
globular clusters and planetary nebula (e.g., Samurovic \& Danziger
2006, Wu \& Tremaine 2006, Douglas et al., 2007), and to much larger radii using satellite
galaxies or weak gravitational lensing. The comparison of lensing and
X-ray masses made by Mahdavi et al. (2007) for a sample of 18 clusters
yielded $M_X/M_L=1.03\pm0.07$ and $M_X/M_L=0.78\pm0.09$ at radii
corresponding to overdensities of 2500 and 500 respectively. These
results, which characterize cluster properties at radii $\sim$0.3-1.5 Mpc,
are broadly consistent with our findings for the inner few tens of
kpc. When combined with our work, these results suggest that the
contribution of non-thermal components to pressure does not exceed
$\sim$10-20\% over the entire cluster volume, as long as the system
has not recently experienced a major merger. In numerical simulations
of structure formation there is a substantial (in the range from a
few per cent to tens of
percent) contribution from residual stochastic gas motions to the
pressure (e.g., Evrard 1990; Nagai, Vikhlinin \& Kravtsov 2007),
especially in the outer regions of clusters. If one accounts for this
apparently unavoidable contribution, the remaining room for cosmic
rays and magnetic fields shrinks further, to $\le$10\% over the whole
cluster volume. Note however that the conditions in the cluster
outskirts and core are markedly different and it is unlikely that a
universally applicable limit can be easily derived.

\section{Conclusions}

Using Chandra observations of M87 and NGC 1399 we derive the
gravitational potential profiles of these two galaxies and compare them
with the potentials derived from optical data. Both the X-ray and
optical methods -- one based on the hydrostatic equilibrium of hot gas
and the other on stellar dynamics -- give consistent changes of the
potential to within $\sim$10-20\% over a broad range  in radial
distance. In M87, several obvious wiggles in the X-ray derived
profile can plausibly be attributed to a spherical shock propagating
through the ICM.

These results suggest that the combined contribution of cosmic rays,
magnetic fields and micro-turbulence $\lesssim$ 10-20\% of the gas
thermal energy. These findings are consistent with the current
paradigm of cool cluster cores, based on the assumption that AGN
activity regulates the thermal state of the gas by injecting energy
into the ICM.

We also show that the limit of $\sim$10-20\% on the energy density
in the form of relativistic protons applies not only to the current
state of the gas, but essentially to the entire history of the ICM,
provided that cosmic ray protons evolve adiabatically and their
spatial diffusion is suppressed.

 The comparison of $\varphi_X$ and
$\varphi_{opt}$ is presented here for two objects only. Given the
uncertainties in optical and X-ray modelling and peculiarities of the
considered objects (e.g. notable shock in M87) it is desirable to
extend the analysis to a larger sample. This project is under way
and the results will be reported in subsequent publications.

\section{Acknowledgements} 
We are grateful to the referee for important comments and suggestions.
This work was supported by the DFG grant CH389/3-2, NASA contracts
NAS8-38248, NAS8-01130, NAS8-03060, the program ``Extended objects in the
Universe'' of the Division of Physical Sciences of the RAS, the Chandra Science Center, the
Smithsonian Institution, MPI f\"{u}r Astrophysik, and MPI f\"{u}r
Extraterrestrische Physik. ST acknowledges support from a Humboldt Research
Award.

\appendix
\section{A Note on  Deprojection Techniques}

\label{sec:deproj}
Mathematically the deprojection procedure described in Section
\ref{sec:deprojection} is identical (except for the $1/\eta$ factor)
to the conventional onion-peeling algorithm (e.g., Fabian et al. 1980)
if the number of spherical shells is equal to the number of radial
annuli and a unique and exact solution exists. However our formulation
(i.e. eq. \ref{eq:chi2}) works also for different numbers of
shells/annuli. Typically $n_s<n_a$, so the inversion problem is
overdetermined, but problems with $n_s>n_a$ are also easily tractable
with the additional condition that the solution should have a minimal
norm. Our method also provides an explicit expression for the
statistical uncertainties in the solution.

Compared to XSPEC's \verb#projct# our deprojection method is much faster,
given the large number of radial shells used here, since it avoids
simultaneous fitting of many parameters. The expectation values are the same
for our method, the onion-peel algorithm and XSPEC's \verb#projct# if the
spectral model is the correct one and the assumption of spherical symmetry
holds. If, however, one of the above assumptions is incorrect, none of the
approaches yields a correct result. The \verb#projct# results can be made more
physical by imposing constraints on the parameters (e.g., non-negative fluxes
in every spherical shell), but at the expense of solving a much more
difficult problem of finding the true minimum in a multi-dimensional
space.

One particular application of XSPEC's the onion peeling
\verb#projct# that is sometimes used is fitting one spherical shell at
a time and moving from outside in. This is faster since only one
spectrum is fitted at each step. This procedure requires the
spectral model to be specified in each outer shell to be able to fit
the inner shells, since the contribution of these shells to the inner
shells is evaluated using the spectral model (convolved with the
instrument effective area and resolution). In contrast in our approach
the observed $1/\eta$ spectra are ``subtracted'' from the inner shells
(after appropriate scaling to account for geometry of shells) and the
spectral analysis deferred to the final step of fitting deprojected
spectra. Hence, any incorrect model assumptions are not
propagated into all the interior shells. Therefore, our deprojection
analysis requires only assumptions about the geometry of the cluster.

The bottom line is that all techniques should yield consistent
results if underlying assumptions (e.g. spherical symmetry and spectral
model) are correct. Our depojection procedure is especially simple and
efficient if many radial bins (spherical shells) are used.

\section{X-ray mass and potential of a non-spherical cluster.}
\label{sec:anonsph}

Here we prove the following result, mentioned in Section \ref{sec:nonsph}:
Consider gas that is in hydrostatic equilibrium in a non-spherical
gravitational potential. Assume that (i) the gas density is characterized by a
power law in radius (a large radius limit of the ellipsoidal $\beta$-model), that is,
\begin{eqnarray}
\rho_{gas}(r,\theta,\phi)=f(\theta,\phi)r^{-\gamma},
\label{eq:rhogasapp}
\end{eqnarray}
where $(r,\theta,\phi)$ are the usual spherical coordinates and
$f(\theta,\phi)$ is an arbitrary positive function; and (ii) the gas
is isothermal. Then the gravitational mass within a sphere of given radius and
the difference in gravitational potential between any two radii at the same
$(\theta,\phi)$ are given correctly by an analysis that assumes that the
gravitational potential and the gas density are spherically symmetric. 

To prove this, we write the hydrostatic equilibrium equation for an isothermal
gas,
\begin{eqnarray}
\rho_{gas}(r,\theta,\phi)\propto \exp[-\mu m_p\varphi({\bf r})/kT].
\end{eqnarray}
Equating this to (\ref{eq:rhogasapp}) and taking the log, we have
\begin{eqnarray}
\varphi(r,\theta,\phi)={kT\over\mu m_p}\left[\gamma\log r-\log
  f(\theta,\phi)\right].
\label{eq:phitri}
\end{eqnarray}
The corresponding  density distribution is 
\begin{eqnarray}
\rho(r,\theta,\phi)&=&\frac{1}{4\pi
G}\nabla^2\varphi(r,\theta,\phi)\nonumber \\ &=&\frac{kT}{4\pi G\mu
  m_p}\left[\frac{\gamma}{r^2}-
\nabla^2 \ln f(\theta,\phi) \right].
\label{eq:arho}
\end{eqnarray}
Here we implicitly assume that $f(\theta,\phi)$ corresponds to a physically
sound density distribution, i.e., $\rho(r,\theta,\phi)\ge 0$ in eq. (\ref{eq:arho}).
For this distribution the gravitating mass within a sphere with radius $r$ is:
\begin{eqnarray}
M(R)&=&\int_{|{\bf r}|<R}d{\bf r}\,\rho(r,\theta,\phi)\nonumber \\
&=&\frac{kT}{\mu m_pG}\left[\gamma r - \int_{|{\bf r}|<R} d{\bf r}\,\nabla^2\ln
  f(\theta,\phi)\right] \nonumber \\ &=&\frac{\gamma kT}{\mu m_pG}r,
\label{eq:mtri}
\end{eqnarray}
since from the divergence theorem $\int_{|{\bf r}|<R} d{\bf r}\nabla^2 \ln
f(\theta,\phi) = \oint_{|{\bf r}|=R} {\bf \nabla} \ln f(\theta,\phi)\cdot
d{\bf s} =0$, where the second integration is done over a surface of a sphere with radius $R$. 

If we assume spherical symmetry when analyzing this cluster (with
arbitrary orientation of the main axis relative to the line of sight),
we will of course recover the same power law dependence of the
density, $\rho_X\propto r^{-\gamma}$.
Then the change in the potential derived from X-ray data will be:
\begin{eqnarray}
\Delta\varphi_X&=&\varphi_X(r_2)-\varphi_X(r_1)\nonumber \\ &=&
\frac{kT}{\mu m_p}\ln \rho_X(r_1)/\rho_X(r_2)=\frac{\gamma kT}{\mu m_p}\ln (r_2/r_1).
\end{eqnarray}
This is exactly the change of the true potential along any direction (eq.\
\ref{eq:phitri}). 

The derived mass within radius $r$ will be:
\begin{eqnarray}
M_X(r)=-\frac{r^2}{G}\frac{1}{\rho_X}\frac{dP_X}{dr}=\frac{r^2}{G}\frac{1}{r}\frac{kT}{\mu
m_p}{d\log\rho_X\over dr}=\frac{\gamma kT}{\mu m_pG}r,
\end{eqnarray}
where $P_X=kT\rho_X/\mu m_p$.  This is the correct value of the mass
within the sphere of radius $r$ (eq.\ \ref{eq:mtri}).  Thus, if the
above conditions are met (isothermal gas with a power law density
dependence), the X-ray analysis yields the correct spherically
averaged mass and potential difference along any radius vector,
independent of the orientation of the triaxial potential with respect
to the line of sight. In other words - treating the triaxial
system as spherically symmetric and performing a deprojection
analysis as for a perfectly spherically symmetric cluster yields the
correct answer on the potential/mass. The same is true for the X-ray
analysis applied to any wedge, no matter how narrow it is.


\label{lastpage}
\end{document}